%% file: main.tex
\documentclass[anonymous=false,sigconf,natbib=false,review=false]{acmart}
\usepackage{multirow}
\usepackage{dblfloatfix}
\usepackage{subcaption}
\usepackage{graphicx}
\usepackage{listings}
\usepackage{wrapfig}
\usepackage{tikz}
\usetikzlibrary{positioning}

\usepackage{hyperref}

\definecolor{linkblue}{RGB}{70, 100, 170}

\definecolor{linkblue}{RGB}{70, 100, 170}
\newcommand{\chref}[2]{\href{#1}{\textcolor{linkblue}{#2}}}

\AtBeginDocument{%
  }

\copyrightyear{2026}
\acmYear{2026}
\setcopyright{cc}
\setcctype{by}
\acmConference[SIGIR '26]{Proceedings of the 49th International ACM SIGIR Conference on Research and Development in Information Retrieval}{July 20--24, 2026}{Melbourne, VIC, Australia}
\acmBooktitle{Proceedings of the 49th International ACM SIGIR Conference on Research and Development in Information Retrieval (SIGIR '26), July 20--24, 2026, Melbourne, VIC, Australia}
\acmDOI{10.1145/3805712.3809627}
\acmISBN{979-8-4007-2599-9/2026/07}

\definecolor{midnightgreen}{rgb}{0.0, 0.29, 0.33}

\RequirePackage[
  datamodel=acmdatamodel,
  style=acmnumeric,
  ]{biblatex}

\begin{document}


\title{Agentic Search in the Wild: \\ Intents and Trajectory Dynamics from 14M+ Real Search Requests}

\author{Jingjie Ning}
\authornote{These authors contributed equally to this work.}
\email{jening@cs.cmu.edu}
\affiliation{%
  \institution{Carnegie Mellon University}
  \city{Pittsburgh}
  \state{PA}
  \country{US}
}

\author{João Coelho}
\authornotemark[1]
\email{jmcoelho@andrew.cmu.edu}
\affiliation{%
  \institution{INESC-ID,~Carnegie~Mellon~University}
  \city{Pittsburgh}
  \state{PA}
  \country{US}
}

\author{Yibo Kong}
\authornotemark[1]
\email{yibok@cs.cmu.edu}
\affiliation{%
  \institution{Carnegie Mellon University}
  \city{Pittsburgh}
  \state{PA}
  \country{US}
}

\author{Yunfan Long}
\authornotemark[1]
\email{justinlo@cs.cmu.edu}
\affiliation{%
  \institution{Carnegie Mellon University}
  \city{Pittsburgh}
  \state{PA}
  \country{US}
}
\author{Bruno Martins}
\email{bruno.g.martins@tecnico.ulisboa.pt}
\affiliation{%
  \institution{INESC-ID, Instituto Superior T\'ecnico, University of Lisbon}
  \city{Lisbon}
  \country{Portugal}
}

\author{Jo\~ao Magalh\~aes}
\email{jm.magalhaes@fct.unl.pt}
\affiliation{%
  \institution{NOVA LINCS\\NOVA University Lisbon}
  \city{Caparica}
  \country{Portugal}
}
\author{Jamie Callan}
\email{callan@cs.cmu.edu}
\affiliation{%
  \institution{Carnegie Mellon University}
  \city{Pittsburgh}
  \state{PA}
  \country{US}
}

\author{Chenyan Xiong}
\email{cx@cs.cmu.edu}
\affiliation{%
  \institution{Carnegie Mellon University}
  \city{Pittsburgh}
  \state{PA}
  \country{US}
}

\renewcommand{\shortauthors}{Ning et al.}

\newcommand{\yunfan}[1]{{\color{cyan} [Yunfan: #1]}}

\begin{abstract}

LLM-powered search agents are increasingly being used for multi-step information seeking tasks, yet the IR community lacks empirical understanding of how agentic search sessions unfold and how retrieved evidence is reflected in later queries. This paper presents a large-scale log analysis of agentic search based on 14.44M search requests (3.97M sessions) collected from DeepResearchGym, i.e., an open-source search API accessed by external agentic clients. We sessionize the logs, assign session-level intents and step-wise query-reformulation labels using LLM-based annotation, and propose Context-driven Term Adoption Rate (CTAR) to quantify whether newly introduced query terms are lexically traceable to previously retrieved evidence.
Our analyses reveal distinctive behavioral patterns. First, over 90\% of multi-turn sessions contain at most ten steps, and 89\% of inter-step intervals fall under one minute. Second, behavior varies by intent. Fact-seeking sessions exhibit high repetition that increases over time, while sessions requiring reasoning sustain broader exploration. Third, query reformulations are often traceable to retrieved evidence across steps. On average, 54\% of newly introduced query terms appear in the accumulated evidence context, with additional traceability to earlier steps beyond the most recent retrieval. These findings provide candidate signals for repetition-aware stopping, intent-adaptive retrieval budgeting, and explicit cross-step context tracking. We released the anonymized logs, making them available at a public HuggingFace~\chref{https://huggingface.co/datasets/cx-cmu/deepresearchgym-agentic-search-logs}{repository}.

\end{abstract}




\maketitle

\input{1-intro}

\input{2-related-works}

\input{3-data-setup}

\input{4-methodology}

\input{5-RQ1}

\input{6-RQ2}

\input{7-RQ3}

\input{8-RQ4}

\input{9-discussion}

\input{11-conclusion}

\begin{acks}
The Portuguese researchers were supported by the Portuguese Recovery and Resilience Plan through project C645008882-00000055 (i.e., the Center For Responsible AI), and also by the Fundação para a Ciência e a Tecnologia, I.P. (FCT) under projects UID/50021/2025 (https://doi.org/10.54499/UID/50021/2025), and UID/PRR/50021/2025 (https://doi.org/10.54499/UID/PRR/50021/2025). João Coelho was additionally supported through the Ph.D. scholarship with reference PRT/BD/153683/2021, under the CMU Portugal Program. 
\end{acks}


\clearpage
\input{appendix}

\clearpage
\printbibliography


\end{document}

%% file: 1-intro.tex
\section{Introduction}
\label{sec:intro}

Information retrieval is shifting from human-initiated search into agentic search~\cite{asai2024selfrag,2022_WebGPT,2023_Toolformer,2023_react}, where LLM-powered agents plan and execute multi-step information seeking with retrieval tools. Instead of issuing a single query and consuming a ranked list, an agent may iteratively reformulate queries, retrieve evidence, and issue later queries in response to the returned context. While agent capabilities are increasingly demonstrated on controlled benchmarks~\cite{2025_Beneficial_Reasoning_Behaviors,2023_gaia,webwalker_dataset}, benchmark scores alone do not reveal how agents' queries evolve across steps, or how  context is reflected in later queries.

These questions matter for practical system design. Agents may spend retrieval budget on repetitive or overly narrow reformulations, fail to explore alternative facets, or carry forward little useful context across steps. Understanding session structure can inform query-policy control, and measuring evidence traceability can guide budget allocation and evaluation design. As agents consume results programmatically, leaving no direct trace of what they found useful, logs lack implicit feedback signals, such as the clicks that anchor traditional behavioral inference. This creates a measurement gap. We can observe sequences of submitted queries and returned evidence, but it remains unclear how sessions unfold, how behavior differs by intent, what reformulation moves dominate, and whether later queries are lexically traceable to evidence returned earlier.

\begin{figure}[t]

    \centering
    \includegraphics[width=1\linewidth]{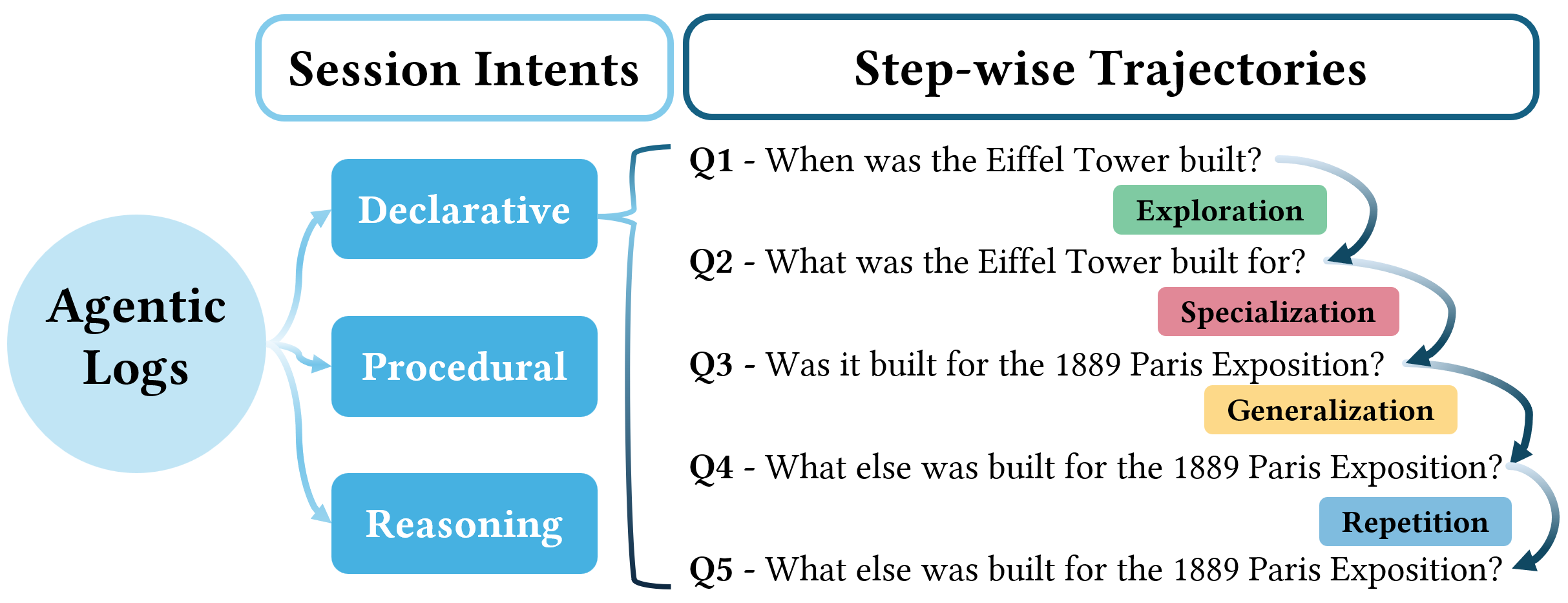}
    \vspace{-8pt}
    \caption{Intent--trajectory structure of agentic search logs.}
    \label{fig:intro_taxonomy}
\end{figure}

To address this gap, we analyze agentic search at two complementary levels, namely what the agent is trying to accomplish in a session, which we capture with session-level intent, and how the agent pursues that goal through step-wise search actions, which we capture with trajectory-level query reformulation. Figure~\ref{fig:intro_taxonomy} illustrates this structure on a short example session. To operationalize these levels, we develop a measurement framework with three components. First, we use LLM-based annotation pipelines to assign interpretable intent and trajectory labels to sessions and step-pairs, following standard taxonomies. Second, we replay logged queries offline to reconstruct the evidence returned at each step. Third, we introduce Context-driven Term Adoption Rate (CTAR), a metric that quantifies whether newly introduced query terms can be lexically traced to retrieved evidence, including traceability to steps beyond the most recent retrieval.

We apply this framework to logs collected from DeepResearchGym (DRGym)~\cite{2025_deep_research_gym}, a research-oriented reproducible search API accessed by external agentic clients. With permission from the DRGym organizers, we study 14.44M logged search requests spanning six months, which we sessionize into 3.97M sessions. The resulting data provides an at-scale view of autonomous agents operating in the wild under a shared retrieval backend, while still preserving the distinction between observable API-level traces and unobserved client-side prompts, memory, and control policies.

Our results characterize retrieval budgeting, reformulation behavior, and evidence traceability in multi-step agentic search. We find that over 90\% of multi-turn sessions contain at most ten steps, and 89\% of inter-step intervals fall under one minute. Retrieval depth, measured by the number of documents requested per query, is largely static, suggesting that many clients treat it as a fixed parameter rather than adapting it within sessions. Behavior also varies by intent. Fact-seeking sessions exhibit the highest repetition, which increases over time, showing that near-duplicate loops can emerge in later steps, while sessions requiring reasoning sustain broader exploration throughout. We also find that many newly introduced query terms are lexically traceable to previously retrieved evidence, with measurable overlap from earlier steps beyond the most recent retrieval.

Our contributions can be summarized as follows:

\begin{itemize}
    \item We provide a large-scale behavioral characterization of agentic search from a reproducible search infrastructure (14.44M search requests, 3.97M sessions), offering an at-scale view of autonomous agents operating in the wild.
    \item We introduce CTAR, as a metric for quantifying evidence-conditioned query evolution, and use it to measure cross-step lexical traceability beyond the most recent retrieval.
    \item We identify candidate design signals from the logs, including repetition-aware stopping, intent-adaptive retrieval budgeting, and cross-step context tracking.
\end{itemize}
\vspace{-0.25em}

We have released the anonymized logs to support future research and reproducibility, making them available as a public HuggingFace dataset \chref{https://huggingface.co/datasets/cx-cmu/deepresearchgym-agentic-search-logs}{repository}.

%% file: 2-related-works.tex
\section{Related Work}

\paragraph{Human Search Behavior and Log Analysis:}
Large-scale query logs have long been used to study search behavior in the wild~\cite{2014_Understanding_User_Behavior,2000_Real_life_information_retrieval,1999_Analysis_of_a_very_large_web_search}, offering scalable, behavior-grounded signals for characterizing session dynamics and query reformulation beyond what offline benchmarks capture.
A core theme is within-session learning. Eickhoff et al.~\cite{2014_Lessons_from_the_journey} trace how newly introduced terms relate to evidence observed before reformulation (e.g., SERP snippets and visited pages), alongside complementary work on interpreting implicit feedback such as clicks and dwell time~\cite{2006_Improving_web_search,2005_Evaluating_implicit_measures,2005_Accurately_interpreting_clickthrough_data}.
Exploratory search and navigation studies further document differences in branching and interaction patterns across users and information needs~\cite{2006_Exploratory_Search,2004_Orienteering,2007_Behavioral_Variability}, while sessionization analyses examine how sessions begin and end in practice~\cite{DBLP:conf/ercimdl/GomesMC19, 2008_Beyond_the_session_timeout, 1999_Analysis_of_a_very_large_web_search}.
We adopt this evidence-traceability perspective for autonomous agents and operationalize it using retrieved evidence text, enabling systematic comparisons of agent behaviors across intents and guiding design choices such as retrieval budgeting and cross-step context management.
While a small line of work compares humans and agents directly~\cite{2025_human_vs_agent,wang2025aiagentshumanwork,zhou2025psychologicalbehaviouralresponseshumanagent}, these comparisons are often task-specific or simulation-based, motivating complementary large-scale log analyses of how autonomous agents search across sessions in the wild.

\paragraph{LLM Interaction Platforms and Usage Logs:}
Recent efforts analyze large-scale interaction data from LLM systems and evaluation platforms.
Chatbot Arena (LMSYS LLM Arena) aggregates pairwise preference votes~\cite{2024_chatbot_arena}, and LMSYS-Chat-1M releases one million multi-model conversations collected in the wild~\cite{2024_lmsys_chat_1m}.
OpenAI reports how people use ChatGPT at scale~\cite{2025_how_people_use_chatgpt}, and Anthropic presents privacy-preserving analyses of millions of Claude conversations to characterize economic task usage~\cite{2025_economic_tasks_ai_claude}.
SciArena extends the Arena protocol to scientific literature-grounded tasks and provides a corresponding benchmark~\cite{2025_scierena}.
These works capture usage and preference signals, but typically do not expose tool-level retrieval traces (queries, evidence, step-wise search decisions) needed to study agentic search behavior and within-session evidence reuse.

\begin{figure*}[!t]
\vspace{-8pt}
    \centering
    \begin{subfigure}[b]{0.33\textwidth}
        \centering
        \includegraphics[width=\textwidth]{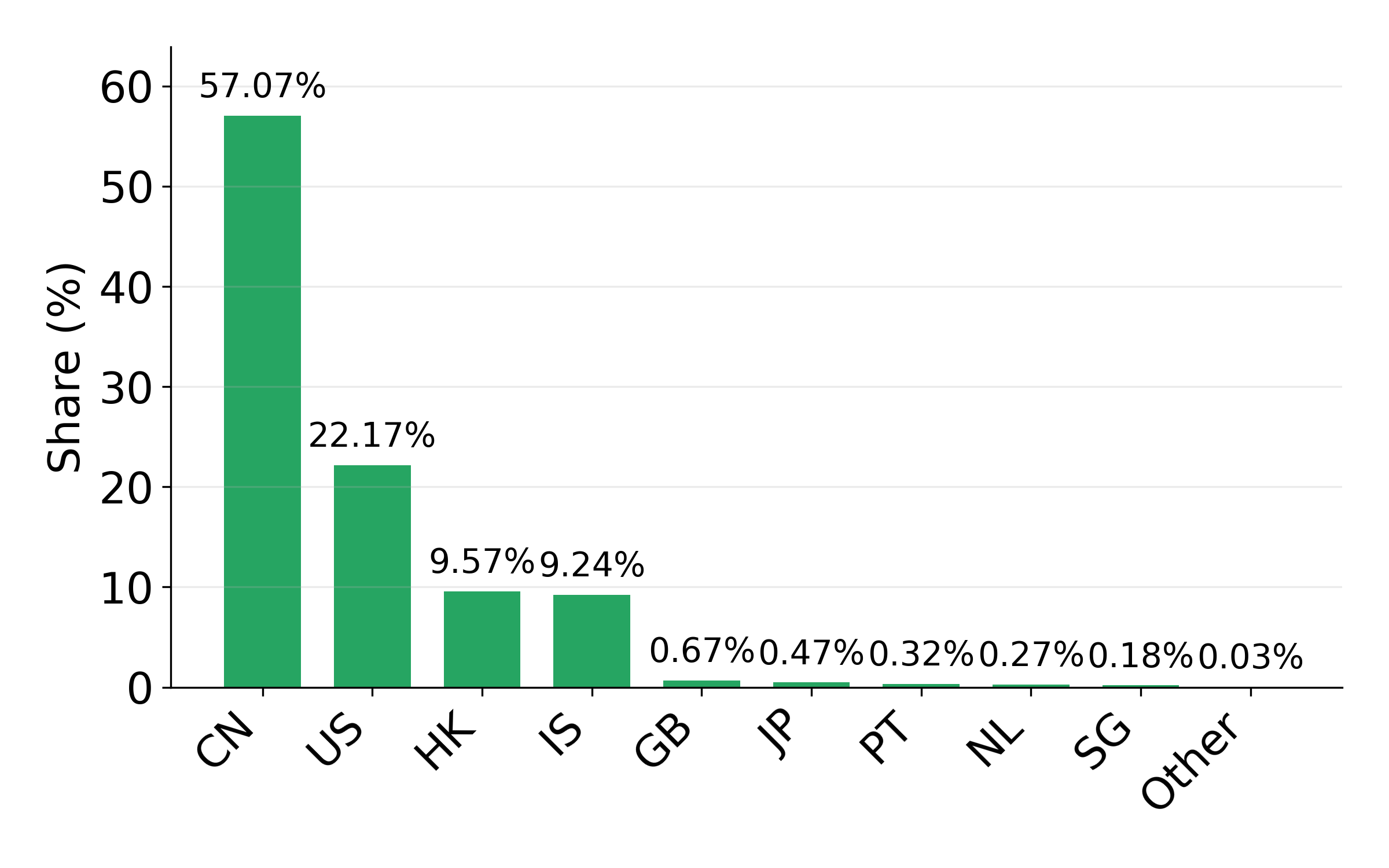}
        \vspace{-6pt}
        \caption{Geographic distribution of requests.}
        \label{fig:basic_country}
    \end{subfigure}
    \hfill
    \begin{subfigure}[b]{0.33\textwidth}
        \centering
        \includegraphics[width=\textwidth]{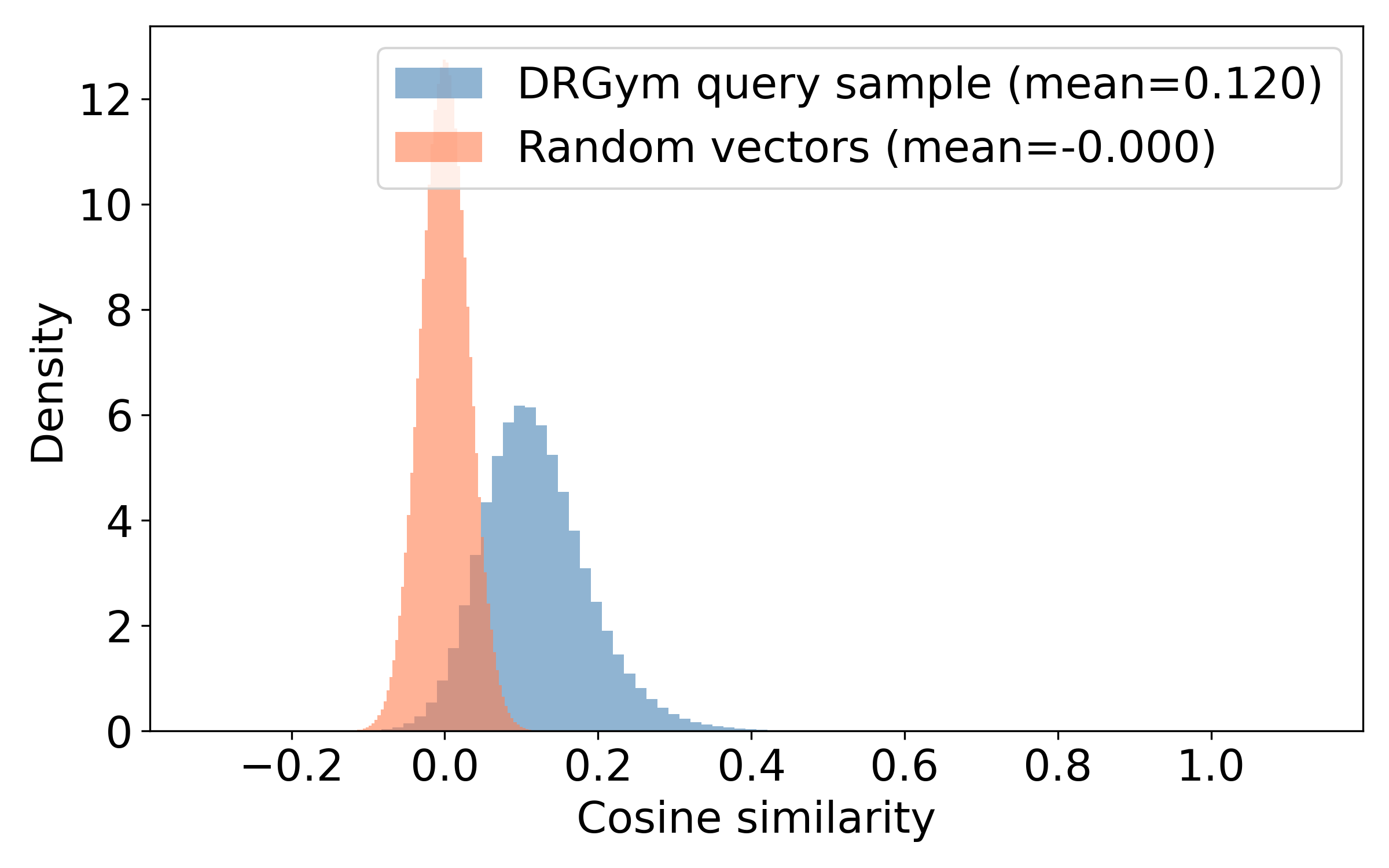}
        \vspace{-6pt}
        \caption{Pairwise query similarity (100k sample).}
        \label{fig:pairwise_dist}
    \end{subfigure}
    \hfill
    \begin{subfigure}[b]{0.33\textwidth}
        \centering
        \includegraphics[width=\textwidth]{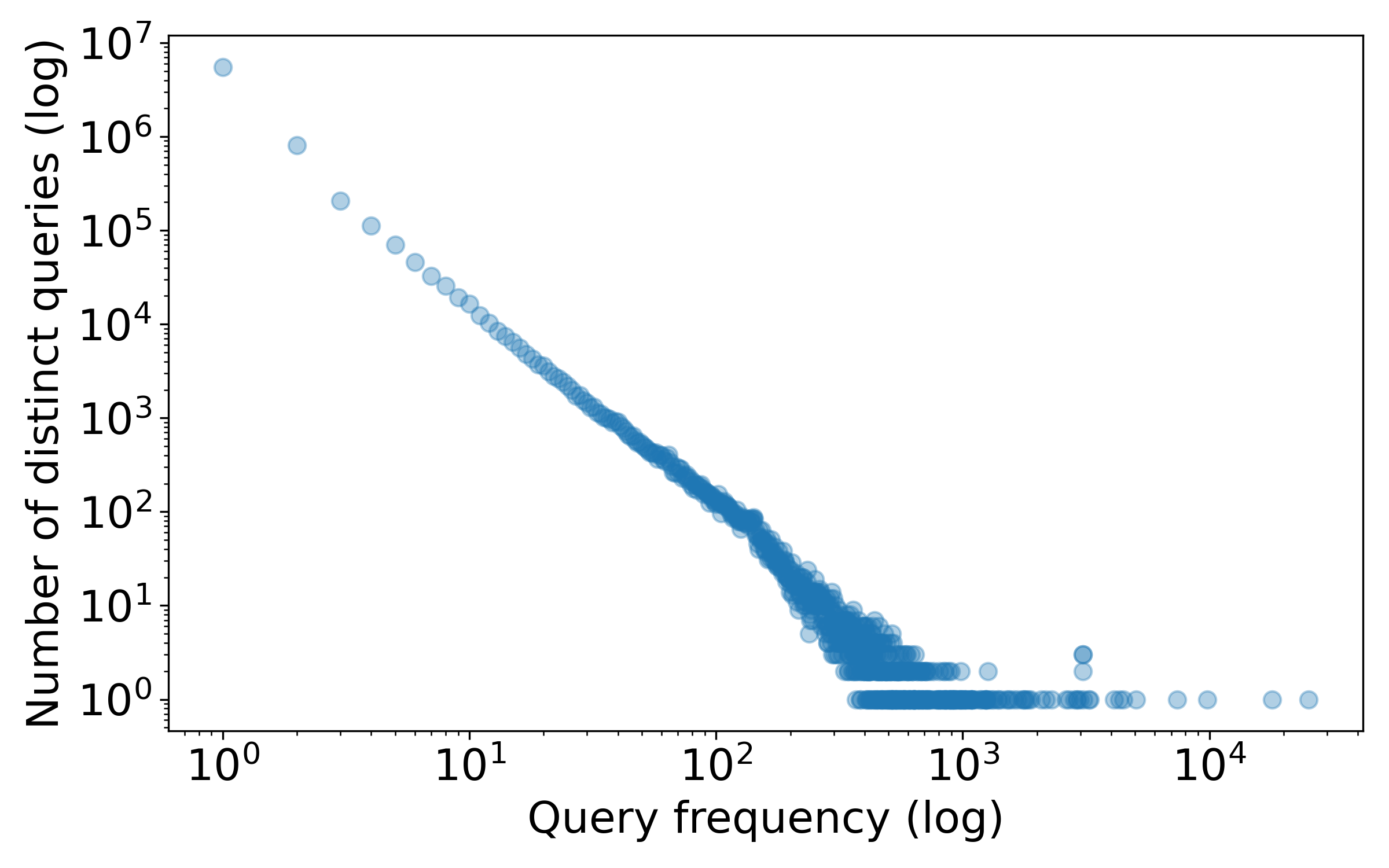}
        \vspace{-6pt}
        \caption{Query frequency distribution (log--log).}
        \label{fig:query_freq_spectrum}
    \end{subfigure}
    \vspace{-8pt}
    \caption{Representativeness and diversity of the DRGym logs.}
\label{fig:repr_div}
\end{figure*}

\paragraph{Agentic Search Modeling, Benchmarks, and Infrastructures:}
Recent systems enabling LLMs to plan multi-step interactions with retrieval tools have shifted IR toward agentic workflows~\cite{asai2024selfrag,2022_WebGPT,2023_Toolformer,2023_react}.
Benchmarks for tool-using agents include WebShop~\cite{2022_webshop}, WebArena~\cite{2023_webarena}, AgentBench~\cite{2023_agentbench}, and large-scale tool-use evaluation such as ToolLLM/ToolBench~\cite{2023_toolllm}.
DeepResearchGym (DRGym) provides an open-source sandbox with a reproducible search API and evaluation protocol for deep research systems~\cite{2025_deep_research_gym}.
Early analyses have begun to formalize agent behaviors. Jin et al.~\cite{2025_Beneficial_Reasoning_Behaviors} link beneficial reasoning patterns to gains on GAIA~\cite{2023_gaia} and WebWalker~\cite{webwalker_dataset}; complementary efforts propose taxonomies and risk frameworks, such as ST-WebAgentBench~\cite{levy2025stwebagentbench} and the Agentic AI Security Scoping Matrix~\cite{2025_aws_agentic_ai_security_scoping_matrix}.
However, benchmark scores alone provide limited visibility into how agents search in practice, and prior human--agent comparisons suggest differences in query breadth and context use that are difficult to diagnose without session-level traces~\cite{2025_human_vs_agent, wang2025aiagentshumanwork, zhou2025psychologicalbehaviouralresponseshumanagent}.
Most prior work focuses on benchmarking and system design, whereas we measure behavior at scale from real logs, and quantify evidence traceability via CTAR.

%% file: 3-data-setup.tex
\section{Data and Log Processing}
\label{sec:data_setup}

In this section, we start by providing an overview of DRGym to better contextualize our analysis. Then, we describe the query log we have been given access to, following with a presentation of the preprocessing and session segmentation (sessionization) pipeline used to convert raw requests into sessions.

\subsection{DRGym Log Overview}
\label{subsec:data_source}
DRGym serves requests from external agentic clients, capturing diverse usage patterns and interaction styles. The API is model-agnostic, i.e., it operates as a retrieval backend rather than a single deployed agent. The logs therefore do not include the client-side model, prompting strategy, memory policy, or the way retrieved evidence is shown to the agent. Requests nevertheless share the same retrieval infrastructure, which lets us study API-level traces, including submitted queries, request parameters, timestamps, and evidence returned by the backend.

The backend performs dense retrieval~\cite{2019_DiskANN,2020_DPR_for_QA} over two large-scale English corpora, i.e. \textit{ClueWeb22-A-EN}~\cite{2022clueweb22} and \textit{FineWeb}~\cite{2024fineweb}. The DRGym paper describes a retrieval API with a \texttt{/search} endpoint for ranked retrieval~\cite{2025_deep_research_gym}. Operating over static web snapshots, rather than a changing live index, enables re-issuing queries under fixed corpora for consistent retrieval behavior across experiments.

Consistent with this design, each log entry records the timestamped query and request parameters (Table~\ref{tab:log_schema}), including retrieval depth and an ANN search-budget parameter~\cite{2019_DiskANN}. For privacy, IP addresses are anonymized and used only for coarse client-level aggregation (grouping, sessionization, and country-level reporting), and never for user identification or fine-grained geolocation.

\paragraph{Scale and Coverage:}
The logs span 2025-06 to 2025-12 and contain 14.44 million requests. After preprocessing and sessionization (Section~\ref{subsec:preprocess_session}), we obtain 3.97M sessions. Requests originate from 558 anonymized client IPs across 25 countries. Figure~\ref{fig:basic_country} shows the geographic coverage, with the largest shares from China and the United States, followed by Hong Kong and Iceland. Overall traffic is substantial, peaking at 2.49 million requests in a single week. 

\begin{table}[t!]
    \vspace{-6pt}
    \small
    \centering
    \caption{Fields recorded in the \texttt{search\_logs} table.}
    \label{tab:log_schema}
    \vspace{-6pt}
    \begin{tabular}{p{0.2\linewidth} p{0.7\linewidth}}
    \toprule
    \textbf{Field} & \textbf{Description} \\
    \midrule
    \texttt{id} & Auto-incremented unique identifier for each request. \\
    \texttt{ip\_address} & Client IP address (anonymized for analysis). \\
    \texttt{query\_text} & Query string submitted to the API. \\
    \texttt{num\_of\_docs} & Number of documents requested for retrieval. \\
    \texttt{complexity} & ANN retrieval complexity controlling search budget. \\
    \texttt{dataset} & Corpus used for retrieval (ClueWeb22 or FineWeb). \\
    \texttt{timestamp} & Timestamp of the logged request. \\
    \bottomrule
    \end{tabular}
    \vspace{-6pt}
\end{table}

\paragraph{Semantic Diversity:}
Figure~\ref{fig:pairwise_dist} plots the pairwise cosine-similarity distribution for 100k randomly sampled queries using Qwen3-Embedding-0.6B~\cite{2025_qwen3embedding}. The distribution (mean$=$0.12) lies close to the random-vector baseline (mean$\approx$0 for uniformly distributed vectors), indicating that queries are semantically diverse rather than clustered around repeated themes. The slight rightward shift likely reflects shared information-seeking phrasing, not semantic redundancy. For reference, Qwen3 uses cosine similarity $>0.7$ to mark semantically related pairs during training~\cite{2025_qwen3embedding}.

\paragraph{Query-Level Repetition:}
Figure~\ref{fig:query_freq_spectrum} shows a long-tailed query frequency distribution. Most distinct queries are quite rare, while only a small set of queries repeats often. In particular, 53.89\% of distinct queries occur at most three times, including 38.38\% singleton queries, and the top-10 and top-100 most frequent queries account for only 0.59\% and 1.51\% of all requests, respectively.

Taken together, the broad geographic coverage, low average semantic similarity, and long-tailed frequency spectrum suggest that the DRGym query stream contains a diverse mixture of information needs, rather than a narrow set of repeatedly executed prompts. 

\begin{table}[t!]
    \small
    \centering
    \caption{Semantic overlap between selected agentic benchmarks and a 1M log query sample (cosine similarity $\geq 0.7$).}
    \label{tab:benchmark_overlap}
    \vspace{-8pt}
    \begin{tabular}{lrrr}
    \toprule
    \textbf{Benchmark} & \textbf{Bench.\ Queries} & \textbf{Hits in Sample} & \textbf{\% of Sample} \\
    \midrule
    GAIA & 103 & 27 & 0.00\% \\
    FRAMES & 824 & 879 & 0.09\% \\
    HLE & 2,158 & 616 & 0.06\% \\
    WebWalkerQA & 680 & 2,219 & 0.22\% \\
    \midrule
    \textit{Total} & 3,765 & 3,741 & 0.37\% \\
    \bottomrule
    \end{tabular}
\end{table}

As an additional sanity check, we estimate whether the stream is concentrated around several widely used agentic and deep-research benchmarks. To that end, we measure semantic overlap between log queries and four widely used benchmarks: GAIA~\cite{2023_gaia}, FRAMES~\cite{frames_dataset}, HLE~\cite{2025_hle}, and WebWalkerQA~\cite{webwalker_dataset}. These benchmarks provide a check against concentration on common deep-research, reasoning, and web navigation tasks. We consider a sample of 1 million queries from the logs, and encode all benchmark queries using Qwen3-Embedding-0.6B. Then, we count log queries exceeding a cosine similarity threshold of 0.7 to any benchmark query. As shown in Table~\ref{tab:benchmark_overlap}, benchmark-similar queries constitute less than 0.4\% of the sample across all four benchmarks combined. Overlap is lowest for GAIA, and highest for WebWalker, whose web traversal queries more closely resemble natural search formulations. Overall, these results suggest that the logs reflect diverse open-ended usage rather than concentrated benchmark execution.

To support reproducibility and enable further research, we have released the cleaned and anonymized logs associated with this study at a public Hugging Face dataset \chref{https://huggingface.co/datasets/cx-cmu/deepresearchgym-agentic-search-logs}{repository}. We have removed direct identifiers (e.g., IP addresses) and applied standard PII scrubbing on free-text fields, releasing only the fields needed to reproduce our analyses with anonymized session IDs. We have documented the anonymization procedure and residual risks in the dataset card, following prior large-scale LLM interaction log releases and privacy-preserving analyses~\cite{2025_economic_tasks_ai_claude,2024_lmsys_chat_1m}. After the initial publication, the dataset may be updated as additional logs are collected and validated.

\subsection{Log Preprocessing and Sessionization}
\label{subsec:preprocess_session}

We first remove malformed entries (e.g., empty queries), internal testing traffic, and outlier repetition bursts, before segmenting the remaining stream into sessions.

Although standard sessionization often relies on fixed time-gap heuristics, agentic requests can arrive in fast parallel patterns~\cite{2025_FlashResearch}, making a pure temporal cutoff unreliable. We therefore sessionize with a semantic-continuity criterion combined with an explicit temporal constraint. Concretely, for each IP we maintain active sessions and assign an incoming query to the most semantically continuous active session when the continuity score exceeds a threshold; otherwise we start a new session. We additionally impose a 10-minute hard cutoff between consecutive queries within a session, reflecting faster interaction loops than the conventional 30-minute rule for human logs~\cite{2008_Beyond_the_session_timeout,1999_Analysis_of_a_very_large_web_search}. Among classifier-predicted continuous pairs, only 0.92\% have gaps exceeding 10 minutes.

The aforementioned pipeline yields 3.97M sessions. Manual spot-checks confirm that the resulting sessions are generally coherent. Full procedural details (i.e., continuity model, thresholds, and validation) are provided in Appendix~\ref{app:sessionization_procedure}.

%% file: 4-methodology.tex
\section{Methodology}
\label{sec:methodology}

\begin{table*}[t!]
    \centering
    \renewcommand{\arraystretch}{1.1}
    \caption{Session-level descriptive statistics by intent type. Formulas for less standard metrics are given in Appendix~\ref{app:metrics}.}
    \label{tab:rq2_desc_stats}
    \begin{tabular}{llccc}
    \hline
    \multirow{2}{*}{\textbf{Statistic}} & \textbf{Metric} & \textbf{Declarative (fact-seeking)} & \textbf{Procedural (how-to)} & \textbf{Reasoning (analytical)} \\ 
     & (\textit{Sample N / Ratio}) & (99.8k / 88.64\%) & (4.5k / 3.96\%) & (8.3k / 7.41\%) \\ \hline
    \multirow{4}{*}{Mean} & Session Length         & 4.03 & 3.81 & 4.03 \\
                          & Retrieval Depth ($K$)  & 7.70 & 37.34 & 24.99 \\
                          & Query Length (whitespace terms)           & 7.59 & 10.58 & 12.69 \\ 
                          & Initial-Final Gap      & 0.21 & 0.22 & 0.28 \\ \hline
    \multirow{2}{*}{Median} & Total Duration (s)   & 40.00 & 26.00 & 31.00 \\
                          & Step Latency (s)       & 17.00 & 13.00 & 14.00 \\ \hline
    \end{tabular}
\end{table*}

To address the questions motivating this work, we require measurements at two levels, namely session-level intent, which captures the type of information need driving a session, and trajectory-level reformulation, which captures how queries change from step to step. We also require a way to assess whether later queries are lexically traceable to evidence returned earlier in the session. For intent and trajectory labeling, we use standard LLM-as-a-judge pipelines~\cite{li2024llmsasjudgescomprehensivesurveyllmbased,2023_llm_as_a_judge}. For evidence traceability, we introduce a new metric, namely the Context-driven Term Adoption Rate (CTAR).

We segment the log into sessions $\mathcal{S}$, where each session $s=(q_1,\ldots,q_{|s|})$ is an ordered sequence of timestamped queries. Retrieval depth is denoted by $K$, corresponding to the logged parameter \texttt{num\_of\_docs}. We analyze behavior at three granularities: global (corpus-wide), session-level (intent-conditioned), and trajectory-level (adjacent query pairs within a session $q_k \rightarrow q_{k+1}$).

\subsection{LLM-based Intent and Trajectory Labeling}
\label{subsec:llm_label_generation}
\paragraph{Session-Level Intent:}
Different information needs may induce different search strategies. For instance, a user seeking a factual answer may behave differently from one debugging a procedure or reasoning through a complex question. To test whether agentic search exhibits such intent-conditioned structure, we label each session with an intent category. We adopt a three-way taxonomy from web search goal modeling~\cite{2002_A_taxonomy_of_web_search,2014_Lessons_from_the_journey,2004_Understanding_user_goals_in_web_search} corresponding to the following classes: Declarative (fact retrieval), Procedural (method execution), and Reasoning (complex synthesis). Since $q_1$ is often already a reformulation, we assign intent from the whole session.

\paragraph{Trajectory-Level Reformulation:}
Intent alone does not reveal how agents iterate within a session. An agent might narrow its query, broaden it, pivot to a related facet, or retry with a near-identical phrasing. These reformulation patterns have implications for retrieval efficiency: excessive repetition wastes budget, while a lack of exploration may leave relevant facets unexamined. To capture these dynamics, we label each adjacent query pair ($q_k \rightarrow q_{k+1}$) with a trajectory type grounded in prior reformulation taxonomies~\cite{2011_Query_reformulation_mining,2009_Analyzing_and_evaluating_query_reformulation}: Specialization (narrowing by adding constraints), Generalization (broadening by relaxing constraints), Exploration (within-topic facet pivots) and Repetition (identical or near-duplicate reformulations). Representative examples are provided in Appendix~\ref{app:examples}.

\paragraph{Implementation}
We implement labeling with \texttt{gpt-5-nano}~\cite{2025_openai_gpt5_nano_model_docs}. We annotate multi-turn sessions with $|s|\!\in\![2,10]$ for intent (one label per session) and all adjacent pairs for trajectories (one label per pair). We focus on this range because our sessionization analysis (Section~\ref{sec:rq1_global_landscape}) reveals that it covers 90.32\% of all multi-turn traffic, representing the core behavior of current agents. For sessions with mixed or ambiguous signals, the judge assigns the dominant intent, and we interpret intent-conditioned results as aggregate trends rather than definitive labels for every individual session. To assess labeling robustness, we compare labels from two models (\texttt{gpt-5-nano} and \texttt{gemini-3-flash-preview}~\cite{2025_gemini_3_flash}) on a 2000-pair random subset, achieving 95.15\% agreement. The remaining disagreements are spread across categories rather than concentrated in any single label. Prompts are provided in Appendix~\ref{app:llm_judge_prompts}. Unless otherwise noted, analyses from Section~6 onward use a random subset of labeled multi-turn sessions under the annotation budget, excluding single-query sessions and outlier long-tail sessions as described in Section~\ref{sec:rq1_global_landscape}. We also compute auxiliary metrics used throughout the paper, each defined at first use with the summary shown in Appendix~\ref{app:metrics}.

\subsection{Context-driven Term Adoption Rate (CTAR)}
\label{subsec:method_ctar}
Agentic search iterates between retrieval and query formulation. Evidence returned at step $k$ may be reflected in later query reformulations. Yet agentic logs provide no direct signal of what the agent actually attended to in retrieved documents, making evidence use hard to observe. We therefore ask a more tractable traceability question: when the agent introduces new query terms at step $k{+}1$, do those terms appear in evidence returned before that step? This aligns with evidence-traceability perspectives in human log analysis~\cite{2014_Lessons_from_the_journey} and searching-as-learning research~\cite{2015_An_Eye_Tracking_Study_of_Query_Reformulation,2016_Towards_Searching_as_Learning,2022_Learning_assessments_in_search_as_learning}, but has not been systematically studied for autonomous agents.

We formulate this idea through Context-driven Term Adoption Rate (CTAR), i.e. the fraction of newly introduced query terms that can be lexically traced to retrieved evidence. We use exact-match tracing rather than semantic similarity because it is interpretable without threshold tuning, robust across domains and query styles, and conservative, i.e., semantic variants would typically yield higher rates by crediting paraphrases and near matches.

Let $Terms(x)$ denote the set of unique, lowercased, and non-stopword tokens that can be extracted from text $x$. For a trajectory $(q_k \rightarrow q_{k+1})$, the set of newly introduced terms is:
\begin{equation}
NewTerms(q_{k+1}, q_k) = Terms(q_{k+1}) \setminus Terms(q_k).
\end{equation}

Let $E_k$ denote the textual evidence returned by the DRGym backend at step $k$. Since raw logs do not store retrieved documents, we reconstruct $E_k$ by querying the DRGym API using the original logged parameters. Because client-side prompts and memory are not logged, $E_k$ should be interpreted as retrievable evidence exposed through the API, rather than a direct observation of what the agent attended to or retained. We consider two context definitions:
\begin{align}
C_k^{\text{last}} &= Terms(E_k), \\
C_k^{\text{agg}}  &= \bigcup_{i=1}^{k} Terms(E_i).
\end{align}
CTAR is the fraction of new terms appearing in the chosen context:
\begin{equation}
\text{CTAR}^{(\cdot)}_k =
\frac{\left| NewTerms(q_{k+1}, q_k) \cap C_k^{(\cdot)} \right|}
{\left| NewTerms(q_{k+1}, q_k) \right|},
\quad (\cdot) \in \{\text{last}, \text{agg}\}.
\end{equation}

In the previous equation, if $NewTerms(q_{k+1}, q_k)=\emptyset$ or $C_k^{(\cdot)}=\emptyset$, we set $\text{CTAR}^{(\cdot)}_k=0$. In summary, CTAR quantifies the degree to which query evolution is lexically traceable to returned evidence. $\text{CTAR}^{last}$ captures traceability to the immediately preceding step, while $\text{CTAR}^{agg}$ captures traceability to evidence returned at any prior step. Comparing the two lets us measure how much additional lexical overlap is introduced by aggregated context, without assuming that the agent causally used or retained that evidence.

\begin{table*}[t!]
    \centering
    \renewcommand{\arraystretch}{1.3}
    \caption{Descriptive statistics by trajectory type. Formulas for less standard metrics are given in Appendix~\ref{app:metrics}.}
    \label{tab:trajectory_stats}
    \begin{tabular}{l@{\hspace{15pt}}l@{\hspace{50pt}}cccc}
    \hline
    \multirow{3}{*}{\textbf{Statistic}} & \textbf{Metric} & \textbf{Specialization} & \textbf{Generalization} & \textbf{Exploration} & \textbf{Repetition} \\ 
                                        & (\textit{Sample N / Ratio}) & (73.8k / 21.76\%) & (30.6k / 9.02\%) & (125.8k / 37.07\%) & (109.1k / 32.15\%) \\
                                        &  & \textit{Narrowing} & \textit{Broadening} & \textit{Facet pivots} & \textit{Near-duplicates} \\ \hline
    \multirow{4}{*}{Mean} & Dense Similarity    & 78.18\% & 80.07\% & 55.08\% & 96.70\% \\
                          & Jaccard Similarity  & 40.13\% & 44.66\% & 25.10\% & 81.63\% \\
                          & Result Overlap      & 22.58\% & 23.47\% & 7.35\%  & 78.23\% \\ 
                          & Delta Query Length  & $+$1.47 & -2.42   & $+$0.04 & -0.08   \\ \hline
    Median                & Step Latency (s)    & 7.00    & 8.00    & 14.00   & 6.00    \\ \hline
    \end{tabular}
\end{table*}

%% file: 5-RQ1.tex
\section{Aggregate Session Statistics}
\label{sec:rq1_global_landscape}
\paragraph{Session Length and Structural Composition}

\begin{figure}[t!]
\vspace{-8pt}
    \centering
    \includegraphics[width=0.50\linewidth]{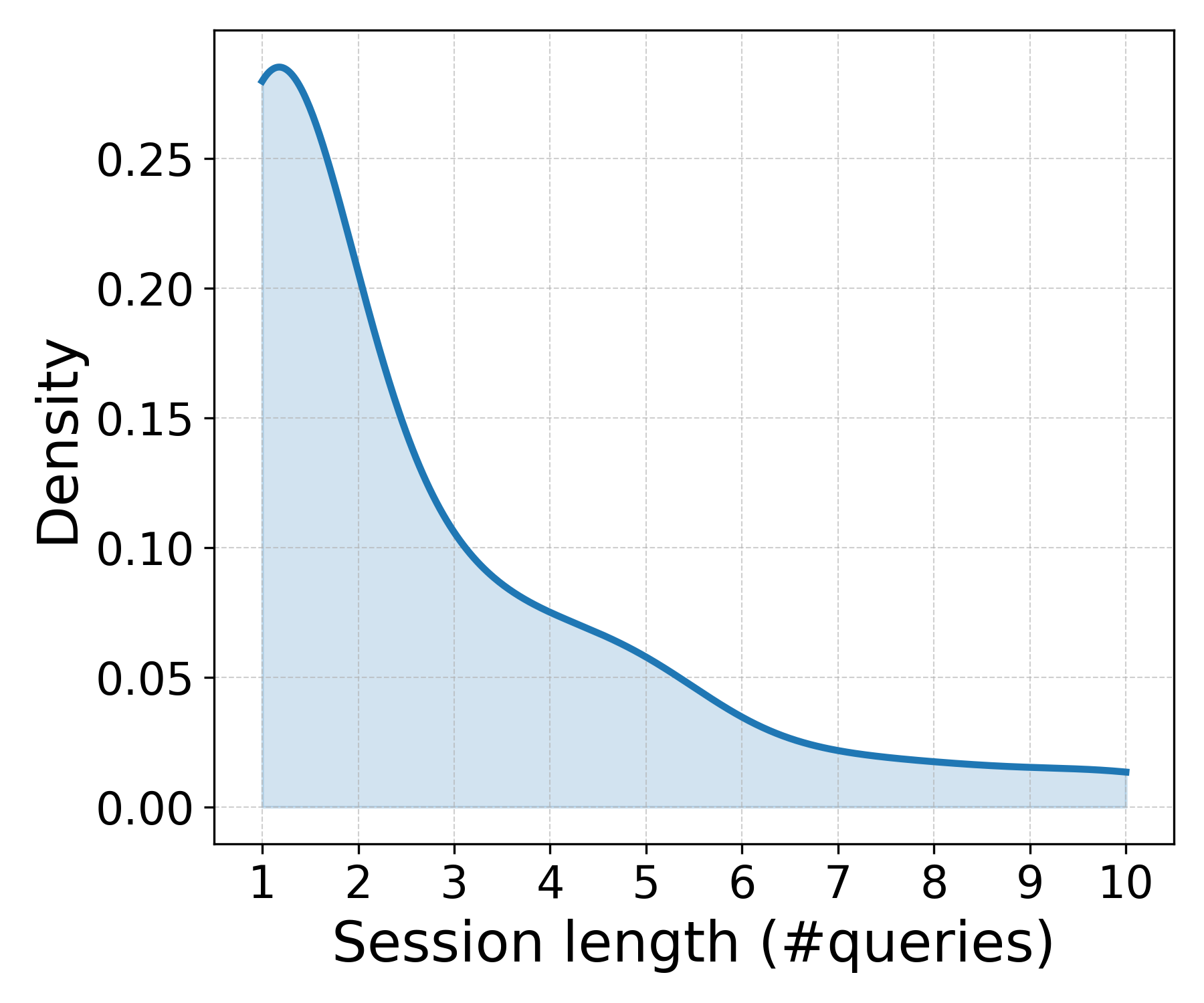}\hfill
    \includegraphics[width=0.50\linewidth]{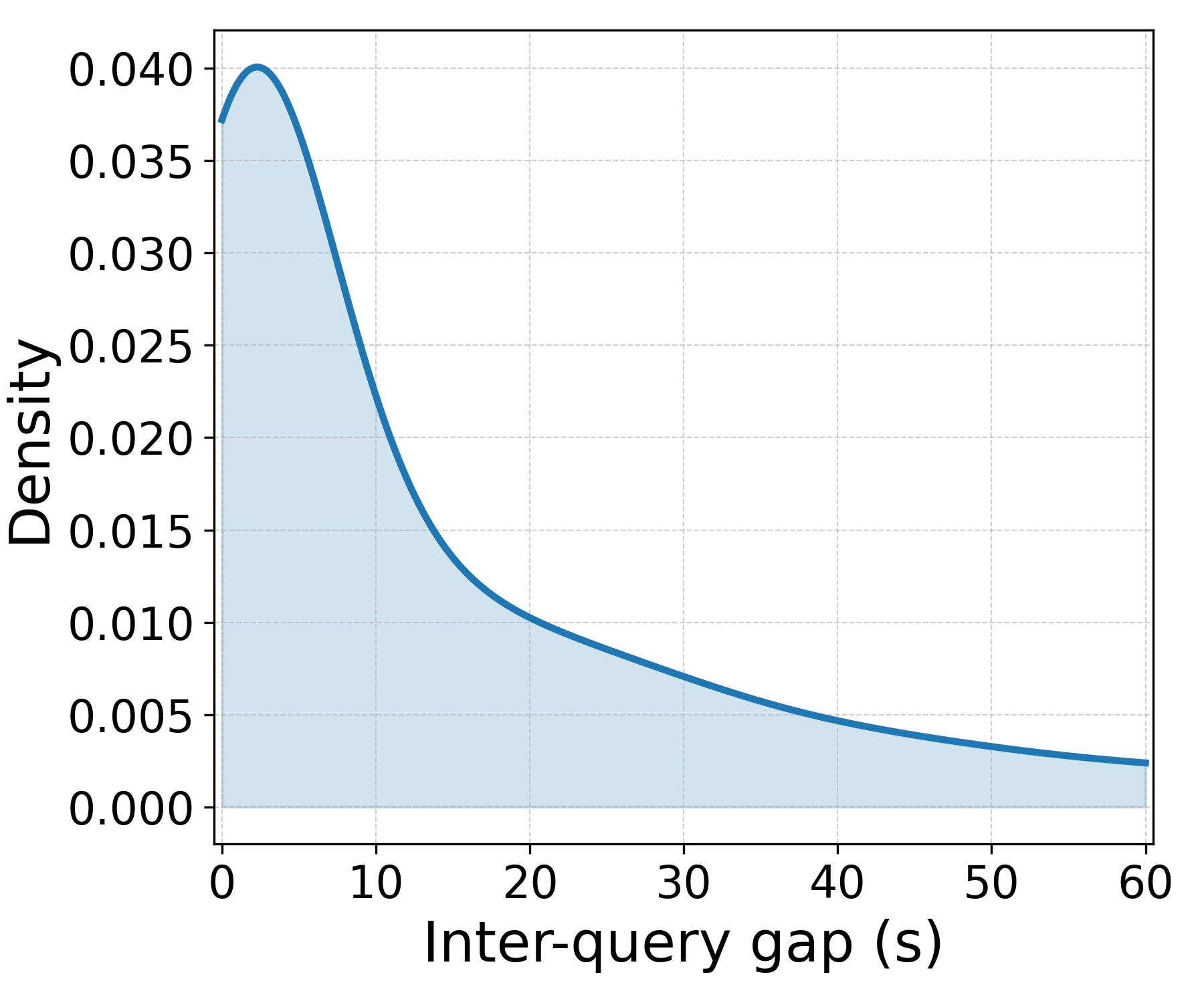}
    \vspace{-8pt}
    \caption{Left: distribution of session length (number of queries per session). Right: distribution of step-wise time intervals between consecutive requests.}
    \label{fig:rq1_kde_dynamics}
    \vspace{-10pt}
\end{figure}

The corpus comprises 3.97M unique search sessions with a skewed length distribution (Figure~\ref{fig:rq1_kde_dynamics}, left). Nearly half (47.77\%) are single-query sessions. Because our trajectory metrics require at least two queries, the rest of the paper focuses on multi-turn sessions. We do not treat single-query sessions as necessarily successful or simple. Instead, our findings characterize sessions that continue beyond one request, which may over-represent complex, uncertain, or iterative information needs. Among multi-turn sessions, 90\% have length $\le 10$, indicating that most multi-turn sessions in the log remain short, though some extend considerably further.

For reference, human web search logs are slightly shorter on average (1.7 queries per session) and include a substantially larger fraction of single-query sessions (77.6\%) \cite{2014_Lessons_from_the_journey,1999_Analysis_of_a_very_large_web_search}. This comparison suggests that agentic systems often engage in more extended information-seeking episodes, but it should not be interpreted as a direct comparison of task success or difficulty. We focus subsequent analyses on multi-turn sessions.

\paragraph{Temporal Dynamics and Interaction Speed}
Within-session intervals are short for most steps, where 56.12\% fall within 0--10 seconds, and 89.21\% are under one minute (Figure~\ref{fig:rq1_kde_dynamics}, right). We use step latency to denote the wall-clock interval between consecutive API requests. This interval may include client-side LLM inference, batching, network delay, and scheduling overhead, so we treat it as a descriptive pacing signal rather than a direct measure of agent deliberation time. Intervals are heavy-tailed, reflecting occasional long-latency steps due to system or pipeline delays.

For reference, prior human log studies report median dwell times of several minutes for knowledge-acquisition intents~\cite{2014_Lessons_from_the_journey}. While dwell time and session duration are not directly comparable to our step latency, the contrast highlights the faster iterative pacing typical of agentic search.

\paragraph{Retrieval Depth and Parameter Stability:}
Retrieval depth is concentrated at fixed values $K\in\{1,5,10\}$, with only 8.36\% of sessions using other values. Furthermore, only 1.35\% of sessions vary $K$ across steps. Since DRGym supports $1 \le K \le 100$, this suggests that many agents treat retrieval count as effectively hard-coded rather than adapted within a session.

%% file: 6-RQ2.tex
\section{Intent-Conditioned Session Behavior}
\label{sec:rq2_intent}
Using the LLM-as-a-judge pipeline described in Section~\ref{sec:methodology}, we label each multi-turn session as either being Declarative (fact-seeking), Procedural (how-to or step-by-step tasks), or Reasoning (comparative, analytical or multi-hop questions). Declarative dominates (88.64\%), followed by Reasoning (7.41\%) and Procedural (3.96\%). Table~\ref{tab:rq2_desc_stats} summarizes the session-level behavior, reporting medians for time-based measures due to heavy tails, and mean values for count and semantic measures.

Beyond length and timing, we characterize sessions with two additional measures. Retrieval Depth summarizes per-step $K$ at the session level, and Initial-Final Gap measures semantic drift from first to last query using $1-\cos(q_1,q_{|s|})$ with Qwen3-Embedding-0.6B~\cite{2025_qwen3embedding}. Formal definitions are given in Appendix B. 

Declarative sessions use the shallowest retrieval yet incur the highest interaction costs. This pattern is consistent with agents using more iterations when per-step retrieval is shallow, although the logs do not determine whether the extra steps improve verification. The pattern contrasts with human fact-finding, where users issue fewer and shorter queries~\cite{2014_Understanding_User_Behavior,2014_Lessons_from_the_journey,1999_Analysis_of_a_very_large_web_search}. Agents instead phrase queries as full constraint-bearing questions, consistent with iterative verification behavior~\cite{2025_Beneficial_Reasoning_Behaviors}.

Procedural sessions show the opposite pattern. Deeper retrieval accompanies a more semantically stable progression, consistent with broader evidence coverage co-occurring with fewer refinement steps. Queries within these sessions are longer than Declarative ones, consistent with prior studies of procedural search that have reported similar characteristics~\cite{2014_Lessons_from_the_journey}.

Finally, we observe that Reasoning sessions match Declarative in turn count but differ in how queries evolve. They show the largest semantic drift and longest queries, while retrieval depth is moderate. The distinguishing signal for Reasoning lies in within-session query reformulation rather than session duration or retrieval depth.

%% file: 7-RQ3.tex
\section{Trajectory Moves and Topologies}
\label{sec:rq3_trajectory}

In this section, we analyze how agents revise queries step by step. This is a defining characteristic of agentic search that exposes intermediate decision-making and supports more fine-grained analysis than single-shot querying. 

\subsection{Trajectory Types and Properties}
\label{subsec:traj_types_properties}

\begin{table}[t!]
    \centering
    \renewcommand{\arraystretch}{1.1}
    \caption{Distribution of trajectory types across all step-wise transitions within sessions of each intent.}
    \label{tab:transposed_trajectory_dist}
    \begin{tabular}{lrrr}
    \hline
     \multirow{2}{*}{\textbf{Trajectory}} & \multicolumn{3}{c}{\textbf{Session Intent}} \\ 
    \cmidrule(lr){2-4}
     & \textbf{Declarative} & \textbf{Procedural} & \textbf{Reasoning} \\ \hline
    Specialization   & 21.12\% & 27.99\% & 26.35\% \\
    Generalization   & 9.07\%  & 10.12\% & 7.89\%  \\
    Exploration      & 36.12\% & 39.04\% & 47.57\% \\
    Repetition       & 33.69\% & 22.85\% & 18.19\% \\ \hline
    \end{tabular}
    \vspace{-8pt}
\end{table}

\begin{figure*}[!tbp]
\vspace{-8pt}
    \centering
    \includegraphics[width=\textwidth]{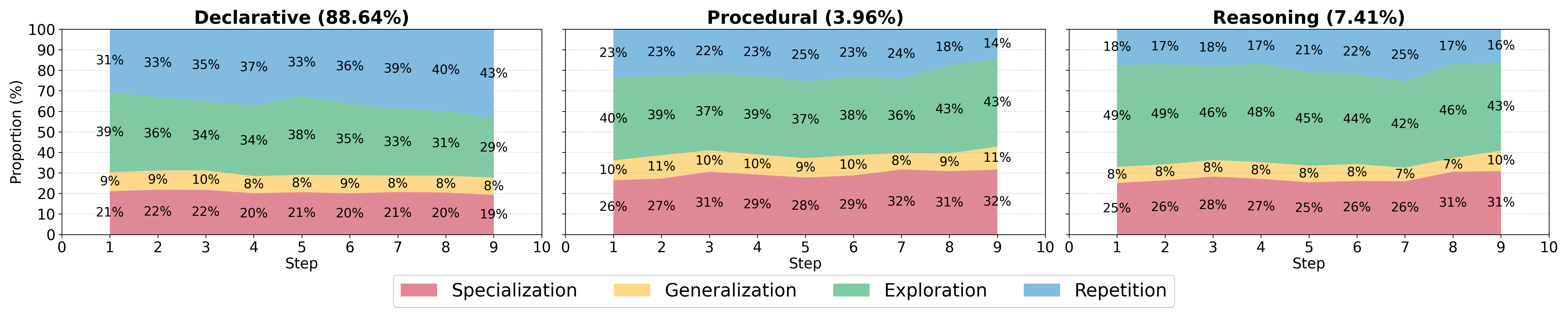}
    \vspace{-8pt}
    \caption{Step-wise trajectory distribution trends for the first 10 steps across different task intents. Each sub-figure illustrates the evolving proportions of Specialization, Generalization, Exploration, and Repetition, as the search session progresses.}
    \label{fig:stepwise_trajectory_evolution}
    \vspace{-8pt}
\end{figure*}

We follow the labeling procedure in Section~\ref{subsec:llm_label_generation}, classifying each adjacent pair $(q_k \rightarrow q_{k+1})$ as Specialization (narrowing by adding constraints), Generalization (broadening by relaxing constraints), Exploration (within-topic facet pivots), or Repetition (identical or near-duplicate reformulations). Table~\ref{tab:trajectory_stats} summarizes trajectory properties, and Table~\ref{tab:transposed_trajectory_dist} reports their usage. We interpret trajectories with three stability measures: \emph{Dense Similarity} is the cosine similarity between consecutive query embeddings (Qwen3-Embedding-0.6B~\cite{2025_qwen3embedding}); \emph{Jaccard Similarity} is the lexical overlap over lowercased, whitespace-tokenized word sets; and \emph{Result Overlap} is the Jaccard overlap between retrieved document identifier sets for consecutive queries (definitions in Appendix~\ref{app:metrics}).

\paragraph{The ``Drill-Down'' Bias:}
Across intents, agents mainly tighten constraints via local edits or pivot across nearby facets, while explicit broadening is consistently the least-used move (under 11\%; Table~\ref{tab:transposed_trajectory_dist}). This imbalance suggests that agents are more comfortable focusing on a local neighborhood of the query space rather than stepping back to relax constraints and reconsider alternatives. Exploration is common (roughly 36--48\%), but pivots tend to induce larger evidence turnover and slower transitions, making them costlier than incremental refinement (Table~\ref{tab:trajectory_stats}). Together, these signatures suggest that agents often continue local edits in high-stability regimes rather than deliberately broadening and re-planning, making such regimes useful targets for future controller analysis.

\paragraph{Intent Differences:}
Although all intents share the same move vocabulary, they exhibit different interaction patterns (Table~\ref{tab:transposed_trajectory_dist}). Declarative sessions are most prone to retry-like behavior, with Repetition at about one-third of moves, consistent with agents re-issuing near-duplicate queries when evidence remains elusive. Reasoning sessions, in contrast, sustain the highest pivoting (Exploration near 48\%) with lower retrying, suggesting a broader search over sub-questions. Procedural sessions more often combine pivots with subsequent constraint tightening, aligning with an ``explore then refine'' workflow in step-by-step tasks.

\begin{figure}[t!]

    \centering
    \hspace*{-1.5em}
    \includegraphics[width=1.0\linewidth]{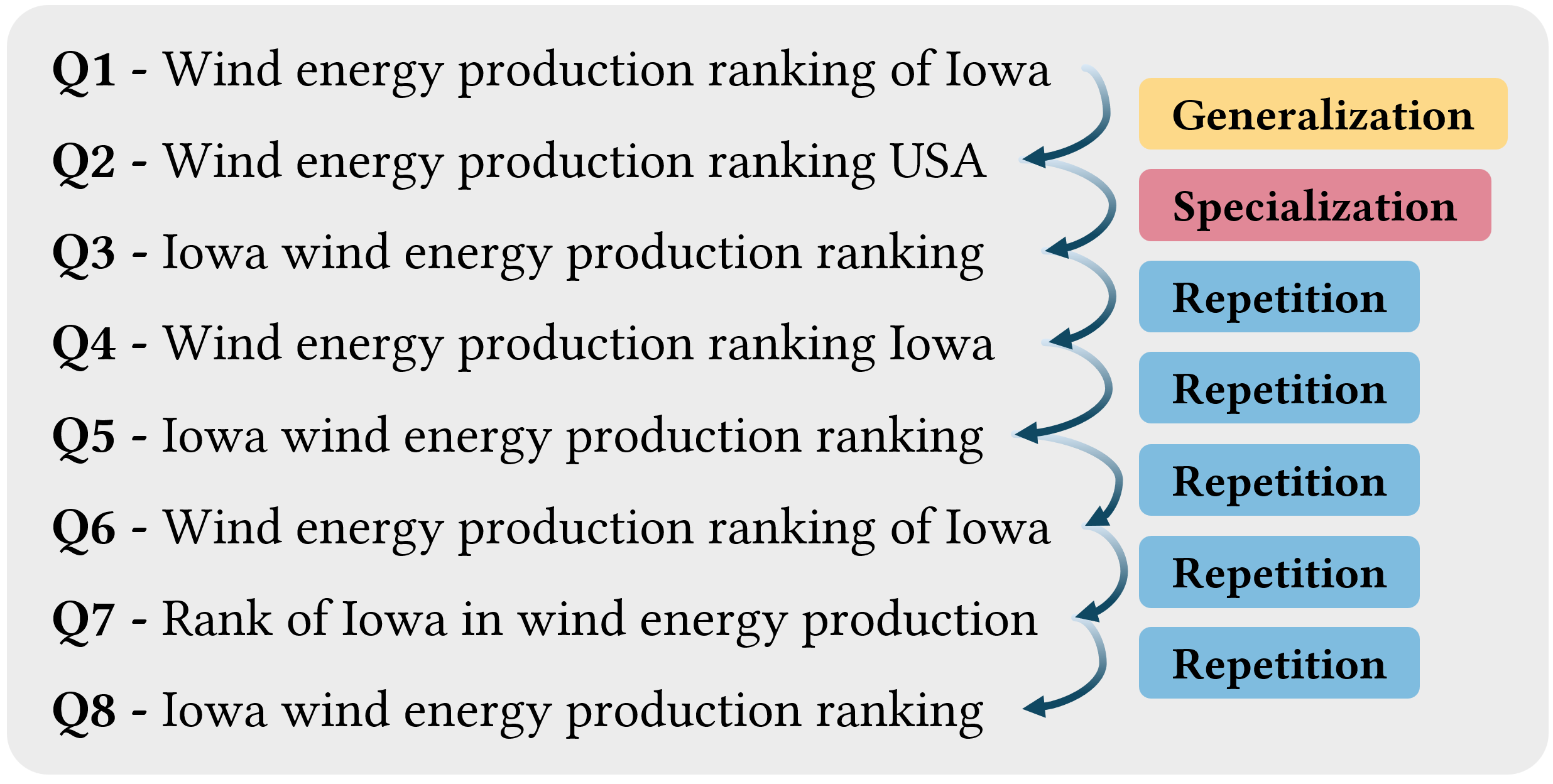}
    \vspace{-8pt}
    \caption{A Declarative retry-loop example dominated by near-duplicate reformulations.}
    \label{fig:case_declarative_loop}
    \vspace{-8pt}
\end{figure}

\paragraph{Stability as a Diagnostic:} 
Table~\ref{tab:trajectory_stats} reveals a stability spectrum. Repetition largely preserves retrieved results (Result Overlap $\sim$78\%), whereas Exploration induces major evidence turnover (Result Overlap $\sim$7\%). Specialization and Generalization fall between these extremes, typically preserving part of the retrieved context while steering the query. This suggests that sustained high-stability runs, especially Repetition, can serve as diagnostic markers of retry-like behavior, whereas Exploration and Specialization more often coincide with evidence turnover. Figure~\ref{fig:case_declarative_loop} illustrates the contrast in a real session. The agent briefly generalizes and then immediately re-specializes (Q1$\to$Q2$\to$Q3), and the interaction then transitions into a high-stability retry loop (Q3--Q8) where the intent remains largely unchanged despite minor wording edits. This example helps explain why Repetition is prominent in Declarative tasks and highlights an intervention point. Detecting such loops could support future tests of strategy-switching policies (e.g., to Exploration) to break the cycle. We connect these regimes to evidence traceability signals, such as term adoption, in Section~\ref{subsec:rq4_ctar}.

\paragraph{Pacing Implications:}
Move types also differ in end-to-end pacing between requests (Table~\ref{tab:trajectory_stats}). Exploration is slower (median of 14.0s) than minor reformulations such as Repetition (6.0s), consistent with facet pivots inducing larger evidence turnover and higher processing cost. This makes strategy selection consequential. When pivots are expensive, agents may default to cheaper local edits, which motivates future tests of when local refinement should be replaced by a different move. Although explicit broadening is rare, it is often followed by re-specialization in the transition dynamics (Section~\ref{subsec:temporal_dynamics}), consistent with broadening acting as a brief reset rather than sustained re-planning.

\begin{figure}[t!]
\vspace{-8pt}
    \centering
    \includegraphics[width=1.0\linewidth]{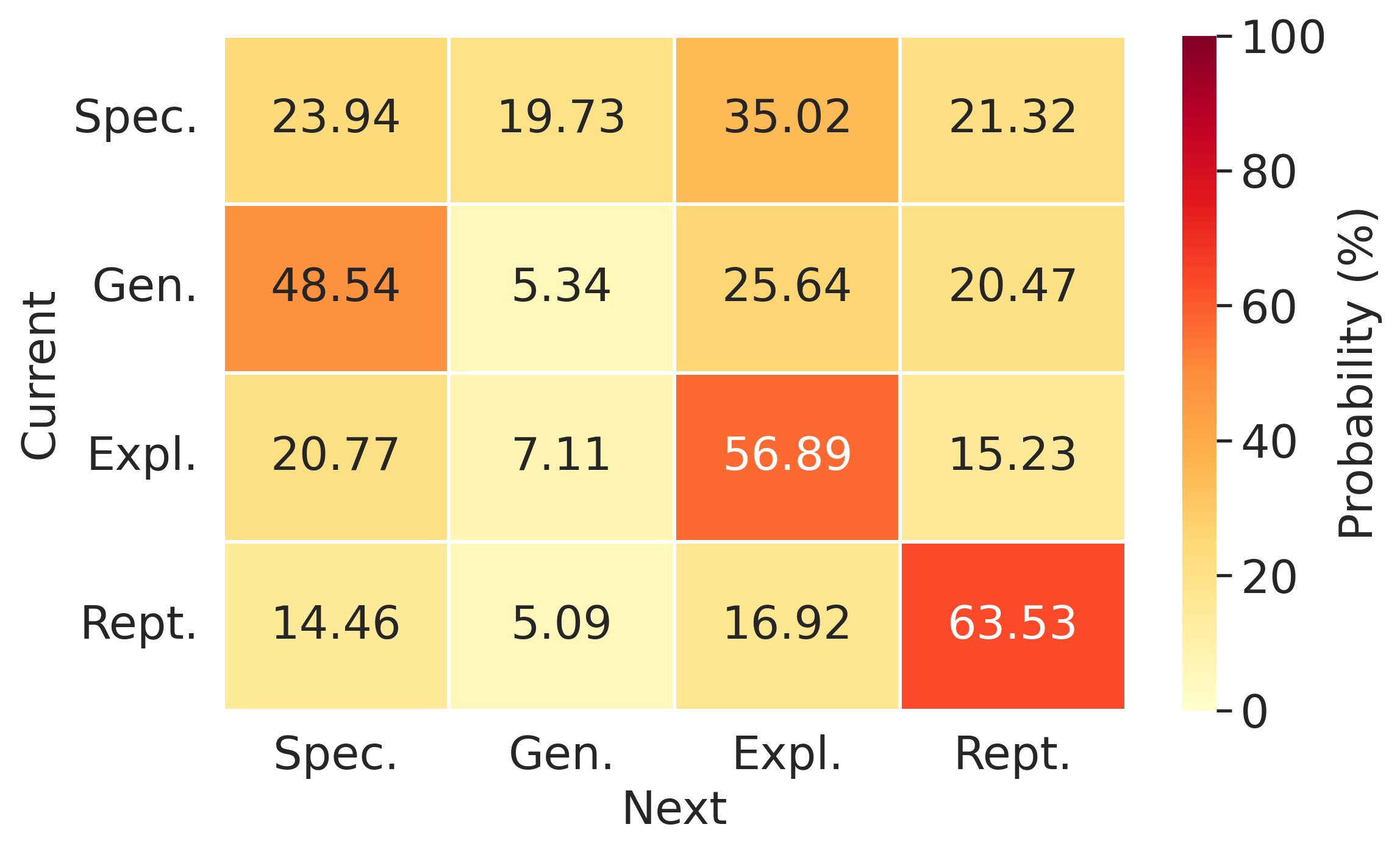}
    \vspace{-8pt}
    \caption{Trajectory transition matrix (row-wise normalized).}
    \label{fig:transition_matrix}
    \vspace{-8pt}
\end{figure}

\begin{figure}[t!]
    \centering
    \hspace*{-1.5em}
    \includegraphics[width=1.0\linewidth]{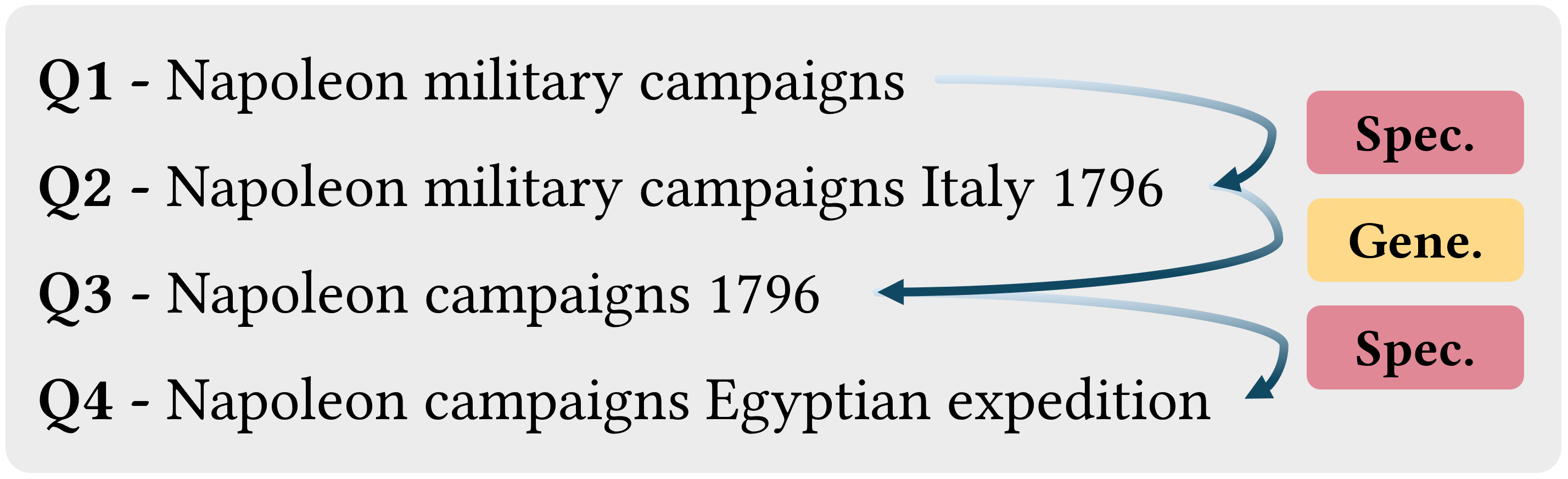}
    \vspace{-8pt}
    \caption{A reset-then-refine example: Specialization $\rightarrow$ Generalization $\rightarrow$ Specialization.}
    \label{fig:case_reset_then_refine}
    \vspace{-8pt}
\end{figure}

\subsection{Temporal Dynamics of Search Strategies}
\label{subsec:temporal_dynamics}

We next study how query-reformulation strategies evolve over a session. Figure~\ref{fig:stepwise_trajectory_evolution}
traces the step-wise trajectory composition over the first 10 steps for each intent, and
Figure~\ref{fig:transition_matrix} summarizes how moves transition from one step to the next.

\paragraph{Trends over Steps:}
Strategy use shifts over time. Early steps mix facet pivots, retries, and constraint adjustments, then diverge by task type (Figure~\ref{fig:stepwise_trajectory_evolution}). Declarative sessions gradually concentrate on retries, consistent with late-stage high-stability behavior; Procedural sessions maintain substantial pivoting but increasingly emphasize refinement; and Reasoning sessions sustain pivoting with consistently low retrying. We report the full step-wise proportions in Figure~\ref{fig:stepwise_trajectory_evolution} and focus here on higher-level directional shifts.

\paragraph{Transition Mechanisms:}
Figure~\ref{fig:transition_matrix} helps explain these trends by showing which moves persist as runs and which act as resets.
Exploration and Repetition often form multi-step runs, consistent with the step-wise patterns in Figure~\ref{fig:stepwise_trajectory_evolution} where pivoting and retrying persist across consecutive steps. In contrast, broadening frequently acts as a brief reset. Nearly half of Generalization moves are followed by Specialization, suggesting that agents relax constraints momentarily before re-introducing them.

\paragraph{Case Study - The ``Reset-then-Refine'' Pattern:}
Figure~\ref{fig:case_reset_then_refine} illustrates a reset-then-refine sequence. The agent specializes a broad topic by adding constraints (Napoleon campaigns $\rightarrow$ Italy 1796), then generalizes by removing them (a shorter, broader query), and re-specializes toward a different facet (Egyptian expedition). Query-length changes match our definitions, where Specialization tends to lengthen queries, while Generalization shortens them (Table~\ref{tab:trajectory_stats}). This is consistent with Generalization acting as lightweight backtracking to switch refinement branches rather than sustained broadening.

%% file: 8-RQ4.tex
\subsection{Context-Driven Term Adoption Rate (CTAR)}
\label{subsec:rq4_ctar}

The previous sections show that agents frequently pivot, refine, and retry across steps, and that these strategies shift over a session's lifespan. A central question is whether such multi-step behavior is traceable to retrieved evidence. Since the logs do not reveal what an agent actually attends to or retains, we study a more limited signal. We test whether new query terms introduced at step $k{+}1$ appear in the evidence returned before that step. We use CTAR as defined in Section.~\ref{subsec:method_ctar} and report both last-step and aggregated variants. 

\paragraph{Lexical Traceability and Recency}
The CTAR analysis indicates that a substantial fraction of newly introduced query terms can be lexically traced to retrieved evidence. Table~\ref{tab:ctar_simplified} reports the mean CTAR under aggregated versus last-step context.

\begin{table}[t!]
    \centering
    \renewcommand{\arraystretch}{1.1}
    \caption{Mean CTAR under Aggregated vs.\ Last-step Evidence.}
    \vspace{-8pt}
    \label{tab:ctar_simplified}
    \begin{tabular}{l@{\hspace{50pt}}c@{\hspace{27pt}}c}
    \hline
    \textbf{Trajectory type} & \textbf{Aggregated} & \textbf{Last-step} \\ \hline
    Specialization             & 78.35\% & 70.93\% \\
    Generalization            & 52.95\% & 49.44\% \\
    Exploration               & 69.59\% & 60.14\% \\
    Repetition                & 20.92\% & 19.77\% \\ \hline
    \textit{Overall}      & 54.35\%            & 48.54\%           \\ \hline
    \end{tabular}
\end{table}

Overall, more than half of newly introduced query terms are present in the aggregated evidence context, with a mean CTAR of 54.35\%. Aggregated context adds 5.81 percentage points over last-step evidence, which is consistent with strong recency effects and some additional lexical traceability to earlier steps~\cite{2023_Lost_in_the_middle}. CTAR also varies substantially by trajectory type. Specialization and Exploration have much higher aggregated CTAR than Repetition, at 78.35\% and 69.59\% compared with 20.92\%. 

CTAR is intentionally a \emph{lexical} traceability measure based on token overlap, rather than a semantic variant, to avoid embedding-model dependence and uninterpretable similarity-threshold choices. It is therefore best viewed as a conservative audit signal. A low CTAR score does not necessarily imply that the agent ignores evidence, since semantic paraphrases and abstractions are not counted. Conversely, a high CTAR score shows that new terms are explicitly present in the returned context, but it does not by itself establish that the evidence caused the reformulation.

\begin{figure}[t!]
    \centering
    \includegraphics[width=0.9\linewidth]{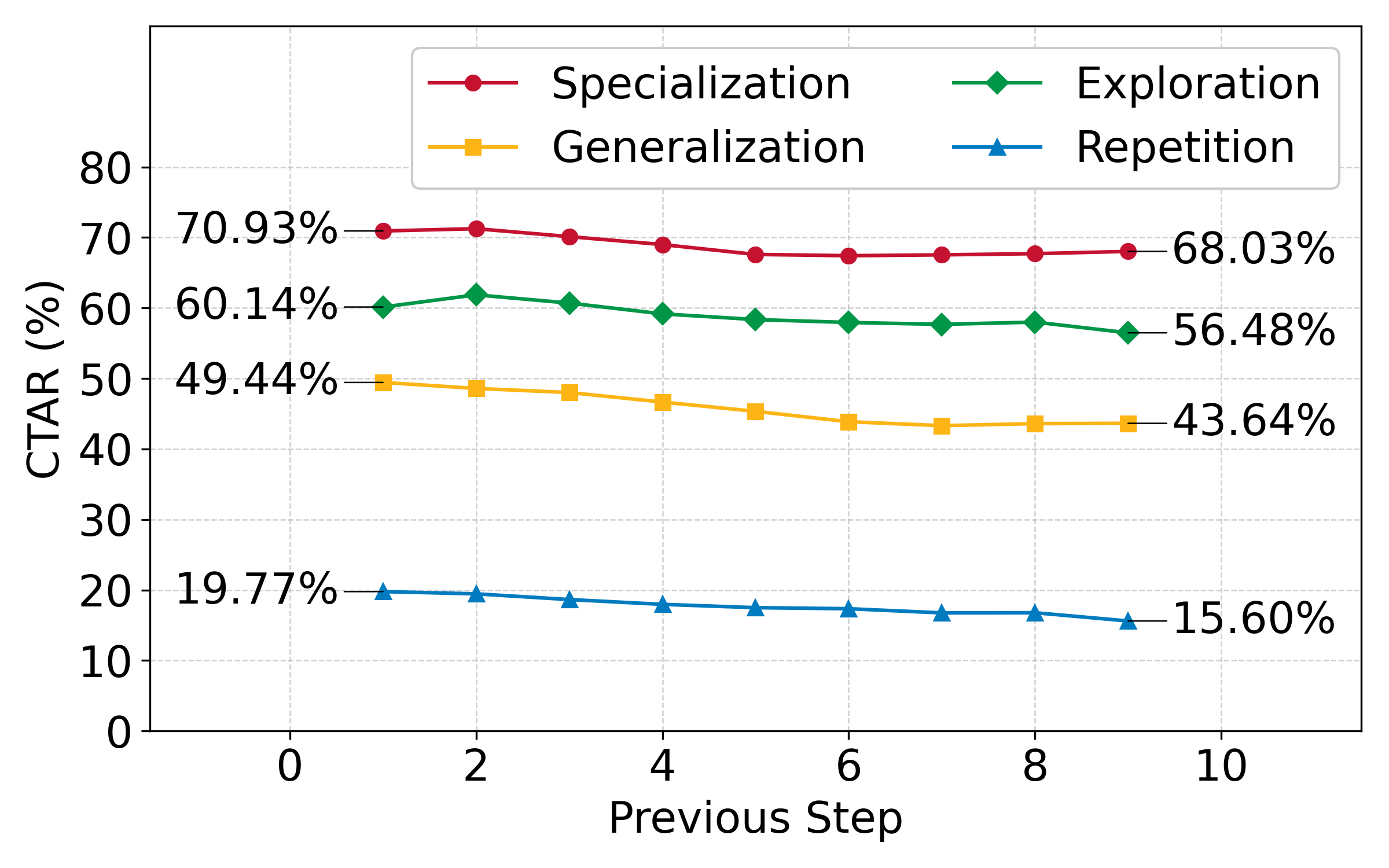}
    \vspace{-8pt}
    \caption{CTAR scores across previous steps.}
    \label{fig:ctar_history}
    \vspace{-8pt}
\end{figure}

\paragraph{Multi-step Context Traceability}
We further examine how CTAR varies when tracing new terms against evidence from different historical steps within a session. Figure~\ref{fig:ctar_history} summarizes the CTAR measured against progressively older evidence contexts. For example, a value of 70.93\% for Spec at ``previous step 1'' means that, for Specialization transitions, 70.93\% of newly introduced terms in $q_{k+1}$ also appear in the evidence retrieved at the immediately preceding step $E_k$, corresponding to the Last Step context in Table~\ref{tab:ctar_simplified}. ``Previous step 2'' analogously traces against $E_{k-1}$, and so on. The curves show that new terms are also lexically traceable to older retrieved evidence. However, because retrieved evidence can overlap across steps, this analysis does not isolate the unique contribution of older evidence. We interpret Figure~\ref{fig:ctar_history} as age-specific traceability, not as proof that the agent directly used earlier documents.

Taken together, these results show that query reformulation is often consistent with evidence-traceable term adoption. New query terms are frequently present in retrieved context, with the strongest signal from the most recent step and weaker but non-trivial traceability to earlier steps. This supports using CTAR as a lightweight diagnostic for cross-step evidence use, while leaving causal interpretation and semantic evidence use to future work.

%% file: 9-discussion.tex
\section{Discussion, Implications, and Limitations}
\label{sec:discussion}

Our analysis provides an observational view of how agentic clients use a shared retrieval backend. The patterns below should be read as diagnostic signals and hypotheses for future systems, rather than evidence that a behavior improves or harms downstream answer quality. The logs do not include success labels such as answer correctness, task completion, or user satisfaction, and they do not expose client-side prompts, memory, or control policies.

\paragraph{Repetition as a Candidate Stall Signal}
In Declarative sessions, repetition increases to 42.68\% by Step~9 (Figure~\ref{fig:stepwise_trajectory_evolution}). This pattern is consistent with near-duplicate loops, although the logs cannot determine whether repetition reflects stalled search, verification, cautious evidence checking, or client-side constraints. Repetition can therefore serve as a candidate signal for future controller policies. Such policies could test whether sustained lexical overlap should lead to a broader query, a different reformulation move, or human review \cite{feild2010predicting,2007_Information_re_retrieval}.

\paragraph{Intent-Adaptive Resource Allocation}
Retrieval depth is largely rigid, with 91.64\% of requests using $K \in \{1,5,10\}$, despite intent-dependent differences in observed usage, such as deeper retrieval for Procedural sessions than Declarative sessions. This suggests that many clients treat retrieval depth as a static parameter. Future architectures could evaluate intent-aware budgeting policies that adjust compute and retrieval depth across intents and steps, rather than relying on fixed $K$ choices \cite{2024_Adaptive_RAG}.

\paragraph{Evidence Grounding as an Audit Signal}
The gain in CTAR from aggregated context (+5.81 pp over the last-step context) is consistent with cross-step lexical traceability, but it does not show that agents actively synthesize or attend to historical evidence. Similarly, the contrast between Specialization (78.35\% CTAR) and Repetition (20.92\% CTAR) shows that lower lexical evidence adoption co-occurs with retry-like transitions, not that low CTAR causes retries. CTAR can therefore be used as a lightweight audit signal for future context management modules that cache prior evidence and surface useful terms for later query formulation \cite{2023_IR_COT}.

%% file: 11-conclusion.tex
\section{Conclusions}
\label{sec:conclusion}

We study agentic search behavior \emph{in the wild} through 14.44M DRGym requests~\cite{2025_deep_research_gym}, converting raw API logs into sessions and analyzing session-level structure, step-wise query transitions, and evidence traceability. Our central takeaway is that multi-step agentic search exhibits measurable intent-conditioned reformulation patterns, even when only API-level traces are available. The diversity analyses in Section~\ref{sec:data_setup} show that the logs are not dominated by a small set of repeated prompts or by the selected benchmark tasks, supporting their use for large-scale behavioral analysis. At the aggregate level, most multi-turn sessions remain short, while retrieval depth is often fixed within a session. This suggests that many current agentic clients rely on repeated query reformulation more than adaptive control of retrieval parameters.

The intent-conditioned and trajectory-level analyses further show that agentic search is not a uniform iterative process. Declarative sessions exhibit more retry-like behavior, while Procedural and Reasoning sessions show different mixtures of refinement, exploration, and repetition. Across trajectories, agents show a drill-down bias, favoring local refinement and facet pivots over deliberate broadening or backtracking. The transition patterns and case studies also show that high-stability runs can emerge within sessions, especially in the context of Declarative tasks. These patterns should not be read as direct evidence of agent success or failure, since the logs do not contain downstream outcome labels. They do, however, provide useful diagnostic structure for studying when an agent continues local edits, when it pivots to a new facet, and when it returns to near-duplicate queries.

To study evidence traceability without clicks or client-side attention signals, we introduce CTAR and measure whether newly introduced query terms appear in retrieved context from earlier steps. We find that many new query terms are lexically traceable to returned evidence, with the strongest signal from the most recent retrieval and weaker but non-trivial traceability to earlier steps. This finding does not establish that agents causally used or retained those documents, nor does it capture semantic paraphrases or abstractions. Instead, CTAR provides a lightweight lexical audit signal for whether query reformulation is consistent with evidence returned by the retrieval backend.

Taken together, the obtained results point to several directions for designing and evaluating agentic IR systems. The move distributions, transition dynamics, and case studies in Section~\ref{sec:rq3_trajectory} suggest that high-stability loops may be useful signals for future controller policies. Such policies could test when repeated local edits should be followed by broader queries, facet pivots, or adaptive retrieval budgeting. Our intent-conditioned analyses also indicate that a single global reformulation policy may be insufficient, since refinement, exploration, and retrying appear with different frequencies across task types. Finally, CTAR can inform future memory and context-management designs by exposing whether query terms remain traceable to prior evidence, while leaving causal evidence use and downstream utility to future evaluation.

Beyond releasing the DRGym-derived dataset and the analysis protocol, future work can connect reformulation moves to downstream answer quality, study which forms of backtracking and evidence reuse are beneficial, and evaluate whether controller interventions improve efficiency or answer quality. More broadly, we hope these measurements and case-driven diagnostics provide a foundation for intent-aware analysis and control of agentic search systems under reproducible retrieval settings.

%% file: appendix.tex
\appendix

\section{Log Sessionization Procedure}
\label{app:sessionization_procedure}
\small

This section describes our log sessionization pipeline, including the semantic continuity model used to link adjacent queries and the per-IP online assignment rules used to form sessions.

\medskip
\noindent\textbf{Step 1: Train a semantic continuity model.}\\
\textbf{(1) Training pairs:} We randomly sample $\sim$200K queries and pair each query with the nearest-in-time query from the same IP. \\
\textbf{(2) Pair labels:} We label each pair as \emph{same-session} vs.\ \emph{different-session} using an LLM-as-a-judge prompt with \texttt{gpt-5-nano-2025-08-07} (Appendix~\ref{app:prompt_pair}). \\
\textbf{(3) Pair representation:} We encode each query with Qwen3-Embedding-0.6B~\cite{2025_qwen3embedding} and use the resulting embeddings to construct a fixed dense feature vector for each query pair as input to the downstream neural classifier. \\
\textbf{(4) Classifier:} We train a 3-layer MLP to output a continuity score in $[0,1]$. The hidden-layer dimensions are 1024, 512, and 256, and the model achieves a held-out accuracy of 0.9419.

\medskip
\noindent\textbf{Step 2: Sessionize online per IP (with validation).}\\
\textbf{(1) Per-IP assignment:} We process queries in order and maintain active sessions for each IP. For an incoming query $q_t$, we score it against each active session using the session's most recent query and assign $q_t$ to the highest-scoring session if the score $\ge 0.5$; otherwise we start a new session. \\
\textbf{(2) Temporal hard cutoff:} Beside the continuity score, if the gap to the candidate session's last query exceeds 10 minutes, we start a new session.  \\
\textbf{(3) Sanity check:} We manually inspect 100 random sessions for coherence; after excluding four unusually long sessions, the remaining sessions are centered on a single objective.

\section{Auxiliary Metric Definitions}
\label{app:metrics}

This section defines the auxiliary metrics used throughout our analyses and provides their formal notation and formulas for reproducibility (Table~\ref{tab:aux_metrics_defs}).

\setlength{\tabcolsep}{4pt}
\renewcommand{\arraystretch}{1.05}
\paragraph{Notation:}
\small
For a session $s=(q_1,\ldots,q_{|s|})$, let $\mathbf{v}_t$ denote the dense embedding of query $q_t$ (and $\cos(\cdot,\cdot)$ the cosine similarity). $W_t$ is the set of normalized tokens from $q_t$ after lowercasing and stopword-aware tokenization $\mathrm{WS\_tok}(\cdot)$. $D_t$ is the set of retrieved evidence returned for query $q_t$ (at the logged retrieval depth).

\begin{table}[H]
    \footnotesize
    \centering
    \caption{Summary of auxiliary metrics used in our analyses.}
    \label{tab:aux_metrics_defs}
    \begin{tabular}{p{0.2\linewidth} p{0.75\linewidth}}
    \toprule
    \textbf{Metric} & \textbf{Formula} \\
    \midrule
    Initial--Final Gap 
    & $\displaystyle \mathrm{Gap}(s)=1-\cos\!\big(\mathbf{v}_1,\ \mathbf{v}_{|s|}\big)$ \\
    
    Dense Sim. 
    & $\displaystyle \mathrm{DenseSim}(q_t,q_{t+1})=\cos\!\big(\mathbf{v}_t,\mathbf{v}_{t+1}\big)$ \\
    
    Jaccard Sim. 
    & $\displaystyle \mathrm{Jac}(q_t,q_{t+1})=\frac{|W_t\cap W_{t+1}|}{|W_t\cup W_{t+1}|},\quad W_t=\mathrm{WS\_tok}(\mathrm{lower}(q_t))$ \\
    
    Result Overlap 
    & $\displaystyle \mathrm{Overlap}(q_t,q_{t+1})=\frac{|D_t\cap D_{t+1}|}{|D_t\cup D_{t+1}|}$ \\
    \bottomrule
    \end{tabular}
\end{table}

\section{Representative Query Examples}
\label{app:examples}
This section presents real representative query examples from our logs to help interpret our intent (Table~\ref{tab:example_queries}) and trajectory (Table~\ref{tab:trajectory_examples}) labels.

\begin{table}[H]
    \footnotesize
    \centering
    \renewcommand{\arraystretch}{1.1}
    \caption{Representative Queries for Intent Categories.}
    \label{tab:example_queries}
    \begin{tabular}{l@{\hspace{20pt}}p{6cm}}
    \hline
    \textbf{Intent} & \textbf{Example Queries} \\ \hline
                & 1. Who owns Handi-Snacks? \\
    Declarative & 2. Definition of home food store \\
                & 3. Ansel Adams residences in Yosemite National Park \\ \hline
                & 1. Best instructions for homemade reading shelf \\
    Procedural  & 2. Practical driving tips for beginners \\
                & 3. How to change a constitution? \\ \hline
                & 1. Why is gas so expensive? \\
    Reasoning   & 2. Does advertising help or harm us? \\
                & 3. Why are minority rights important? \\ \hline
    \end{tabular}
    \vspace{-6pt}
\end{table}

\begin{table}[H]
    \footnotesize
    \centering
    \renewcommand{\arraystretch}{1.1}
    \caption{Representative Transitions Trajectories. Each entry illustrates a step-wise reformulation ($q_k \rightarrow q_{k+1}$).}
    \label{tab:trajectory_examples}
    \begin{tabular}{l@{\hspace{25pt}}c@{\hspace{25pt}}l}
    \hline
    \textbf{Type} & \textbf{Step} & \textbf{Example Queries} \\ \hline
    \multirow{2}{*}{Specialization} & 1 & Recent climate data Durban \\
                           & 2 & Average temperature in Durban 2025 \\ \hline
    \multirow{2}{*}{Generalization}  & 1 & Key events that ended Hitler's dictatorship \\
                           & 2 & Hitler's dictatorship \\ \hline
    \multirow{2}{*}{Exploration} & 1 & Headquarters of Oberoi Hotels \\
                           & 2 & Parent company of Oberoi Hotels \\ \hline
    \multirow{2}{*}{Repetition} & 1 & Handi-Snacks parent company \\
                           & 2 & Who owns Handi-Snacks \\ \hline
    \end{tabular}
    \vspace{-6pt}
\end{table}

\section{LLM-as-a-judge Prompts and Parsing Details}
\label{app:llm_judge_prompts}

This section provides the exact LLM-as-a-judge prompts for reproducibility.

\subsection{Query-pair Continuity Judgment Prompt}
\label{app:prompt_pair}
\small

\begin{lstlisting}[language={},basicstyle=\ttfamily\tiny,breaklines=true,frame=single,label={lst:prompt_pair}]
[SYSTEM]
You must label query pairs for a DeepResearch search agent. The agent fans out several queries to answer ONE user question.
Answer YES if both queries would naturally be used for the same research task or user question (same core topic), even if they cover different aspects or levels of detail.
Answer NO if the queries clearly correspond to different questions, even if they share broad words like 'WWII', 'health', or 'economy'.

[USER]
Query 1: <<query1>>
Query 2: <<query2>>
For a DeepResearch agent that fans out queries to answer ONE user question,
would these two queries belong to the same research task?
Answer YES or NO only.
\end{lstlisting}

\subsection{Session-level Intent Classification Prompt}
\label{app:prompt_intent}

\begin{lstlisting}[language={},basicstyle=\ttfamily\tiny,breaklines=true,frame=single,label={lst:prompt_intent}]
[SYSTEM]
You are an expert search intent classifier.

[USER]
"Session Queries:\
<<joined_queries>>
Classify the user intent of this session into exactly ONE of these three categories:
1. Declarative: Asking for simple facts, definitions, entity attributes, or lists (e.g., 'who is', 'what is', 'release date').
2. Procedural: Asking for steps, methods, tutorials, or guides (e.g., 'how to', 'guide for', 'fix error').
3. Reasoning: Asking for comparisons, planning, analysis, multi-hop reasoning, or creative generation (e.g., 'difference between', 'best plan for', 'why is').
Output ONLY the category name (Declarative, Procedural, or Reasoning).
\end{lstlisting}

\subsection{Step-wise Trajectory Classification Prompt}
\label{app:prompt_trajectory}

\begin{lstlisting}[language={},basicstyle=\ttfamily\tiny,breaklines=true,frame=single,label={lst:prompt_trajectory}]
[SYSTEM]
You are an expert search behavior analyst.

[USER]
Query 1 (Previous): <<PLACEHOLDER: q_k>>
Query 2 (Current):  <<PLACEHOLDER: q_{k+1}>>
Analyze the search behavior evolution from Query 1 to Query 2 for an autonomous agent.
Classify the transition into exactly ONE of these four categories:
1. Specialization (Vertical Deepening): Query 2 is MORE specific than Query 1 by adding constraints/details (q2 \subset q1). (e.g., 'apple' -> 'green apple nutritional value').
2. Generalization (Vertical Broadening): Query 2 is MORE general than Query 1 by removing constraints/abstracting (q2 \supset q1). (e.g., 'green apple nutritional value' -> 'benefits of fruits').
3. Exploration (Horizontal Expansion within the same domain/task): Query 2 is NOT simply more specific or more general. It shifts to a different aspect/subtopic/related entity but still within the same overall topic/domain. (e.g., 'green apple nutritional value' -> 'green apple recipes' or 'MRI Scans' -> 'CT Scans').
4. Repetition (Stationary): Query 2 is semantically equivalent to Query 1. It is a paraphrase, reformatting, or synonym replacement with NO significant change in intent (e.g., 'green apple value' -> 'nutritional value of green apple').
Output ONLY the category name (Specialization, Generalization, Exploration, or Repetition).
\end{lstlisting}

%% file: reference.bib
@article{1999_Analysis_of_a_very_large_web_search,
author = {Silverstein, Craig and Marais, Hannes and Henzinger, Monika and Moricz, Michael},
title = {Analysis of a very large web search engine query log},
year = {1999},
journal = {SIGIR Forum},
}

@article{2000_Real_life_information_retrieval,
author = {Jansen, Bernard J. and Spink, Amanda and Bateman, Judy and Saracevic, Tefko},
title = {Real life information retrieval: a study of user queries on the Web},
year = {1998},
journal = {SIGIR Forum},
}

@article{2014_Understanding_User_Behavior,
author = {Dumais, Susan and Jeffries, Robin and Russell, Daniel M. and Tang, Diane and Teevan, Jaime},
title = {{Understanding User Behavior Through Log Data and Analysis}},
journal = {Ways of Knowing in HCI},
year = {2014},
}

@inproceedings{2014_Lessons_from_the_journey,
author = {Eickhoff, Carsten and Teevan, Jaime and White, Ryen and Dumais, Susan},
title = {{Lessons from the Journey: A Query Log Analysis of Within-session Learning}},
year = {2014},
booktitle = {International Conference on Web Search and Data Mining (WSDM)}
}

@inproceedings{2005_Accurately_interpreting_clickthrough_data,
author = {Joachims, Thorsten and Granka, Laura and Pan, Bing and Hembrooke, Helene and Gay, Geri},
title = {Accurately interpreting clickthrough data as implicit feedback},
year = {2005},
booktitle = {International Conference on Research and Development in Information Retrieval (SIGIR)}
}

@inproceedings{2006_Improving_web_search,
author = {Agichtein, Eugene and Brill, Eric and Dumais, Susan},
title = {Improving web search ranking by incorporating user behavior information},
year = {2006},
booktitle = {International Conference on Research and Development in Information Retrieval (SIGIR)},
}

@article{2005_Evaluating_implicit_measures,
author = {Fox, Steve and Karnawat, Kuldeep and Mydland, Mark and Dumais, Susan and White, Thomas},
title = {Evaluating implicit measures to improve web search},
year = {2005},
journal = {ACM Transactions on Information Systems},
}

@misc{2022clueweb22,
      title={{ClueWeb22: 10 Billion Web Documents with Visual and Semantic Information}}, 
      author={Arnold Overwijk and Chenyan Xiong and Xiao Liu and Cameron VandenBerg and Jamie Callan},
      year={2022},
      eprint={2211.15848},
      archivePrefix={arXiv},
}

@misc{2024fineweb,
      title={{The FineWeb Datasets: Decanting the Web for the Finest Text Data at Scale}}, 
      author={Guilherme Penedo and Hynek Kydlíček and Loubna Ben allal and Anton Lozhkov and Margaret Mitchell and Colin Raffel and Leandro Von Werra and Thomas Wolf},
      year={2024},
      eprint={2406.17557},
      archivePrefix={arXiv},
}

@misc{2025_Beneficial_Reasoning_Behaviors,
      title={{Beneficial Reasoning Behaviors in Agentic Search and Effective Post-training to Obtain Them}}, 
      author={Jiahe Jin and Abhijay Paladugu and Chenyan Xiong},
      year={2025},
      eprint={2510.06534},
      archivePrefix={arXiv}
}

@misc{2023_gaia,
      title={{GAIA: a benchmark for General AI Assistants}}, 
      author={Grégoire Mialon and Clémentine Fourrier and Craig Swift and Thomas Wolf and Yann LeCun and Thomas Scialom},
      year={2023},
      eprint={2311.12983},
      archivePrefix={arXiv},
}

@misc{2025_human_vs_agent,
      title={{Human vs. Agent in Task-Oriented Conversations}}, 
      author={Zhefan Wang and Ning Geng and Zhiqiang Guo and Weizhi Ma and Min Zhang},
      year={2025},
      eprint={2509.17619},
      archivePrefix={arXiv},
}

@misc{2020_DPR_for_QA,
      title={{Dense Passage Retrieval for Open-Domain Question Answering}}, 
      author={Vladimir Karpukhin and Barlas Oğuz and Sewon Min and Patrick Lewis and Ledell Wu and Sergey Edunov and Danqi Chen and Wen-tau Yih},
      year={2020},
      eprint={2004.04906},
      archivePrefix={arXiv},
}

@inproceedings{2019_DiskANN,
 author = {Jayaram Subramanya, Suhas and Devvrit, Fnu and Simhadri, Harsha Vardhan and Krishnawamy, Ravishankar and Kadekodi, Rohan},
 booktitle = {International Conference on Neural Information Processing Systems (NeurIPS)},
 title = {{DiskANN: Fast Accurate Billion-point Nearest Neighbor Search on a Single Node}},
 year = {2019}
}

@inproceedings{2008_Beyond_the_session_timeout,
author = {Jones, Rosie and Klinkner, Kristina Lisa},
title = {Beyond the session timeout: automatic hierarchical segmentation of search topics in query logs},
year = {2008},
booktitle = {Conference on Information and Knowledge Management (CIKM)},
}

@inproceedings{2023_llm_as_a_judge,
author = {Zheng, Lianmin and Chiang, Wei-Lin and Sheng, Ying and Zhuang, Siyuan and Wu, Zhanghao and Zhuang, Yonghao and Lin, Zi and Li, Zhuohan and Li, Dacheng and Xing, Eric P. and Zhang, Hao and Gonzalez, Joseph E. and Stoica, Ion},
title = {{Judging LLM-as-a-judge with MT-bench and Chatbot Arena}},
year = {2023},
booktitle = {International Conference on Neural Information Processing Systems (NeurIPS)},
}

@misc{2023_react,
      title={{ReAct: Synergizing Reasoning and Acting in Language Models}}, 
      author={Shunyu Yao and Jeffrey Zhao and Dian Yu and Nan Du and Izhak Shafran and Karthik Narasimhan and Yuan Cao},
      year={2023},
      eprint={2210.03629},
      archivePrefix={arXiv},
}

@misc{2022_WebGPT,
      title={{WebGPT: Browser-assisted question-answering with human feedback}}, 
      author={Reiichiro Nakano and Jacob Hilton and Suchir Balaji and Jeff Wu and Long Ouyang and Christina Kim and Christopher Hesse and Shantanu Jain and Vineet Kosaraju and William Saunders and Xu Jiang and Karl Cobbe and Tyna Eloundou and Gretchen Krueger and Kevin Button and Matthew Knight and Benjamin Chess and John Schulman},
      year={2022},
      eprint={2112.09332},
      archivePrefix={arXiv},
}

@misc{2023_Toolformer,
      title={{Toolformer: Language Models Can Teach Themselves to Use Tools}}, 
      author={Timo Schick and Jane Dwivedi-Yu and Roberto Dessì and Roberta Raileanu and Maria Lomeli and Luke Zettlemoyer and Nicola Cancedda and Thomas Scialom},
      year={2023},
      eprint={2302.04761},
      archivePrefix={arXiv},
}

@article{2006_Exploratory_Search,
author = {Marchionini, Gary},
title = {Exploratory search: from finding to understanding},
year = {2006},
journal = {Communications of the ACM},
}

@inproceedings{2004_Orienteering,
author = {Teevan, Jaime and Alvarado, Christine and Ackerman, Mark S. and Karger, David R.},
title = {The perfect search engine is not enough: a study of orienteering behavior in directed search},
year = {2004},
booktitle = {Conference on Human Factors in Computing Systems (CHI)},
}

@inproceedings{2007_Behavioral_Variability,
author = {White, Ryen W. and Drucker, Steven M.},
title = {Investigating Behavioral Variability in Web Search},
year = {2007},
booktitle = {International Conference on World Wide Web (WWW)},
}

@misc{2025_aws_agentic_ai_security_scoping_matrix,
  author       = {Aaron Brown and Matt Saner},
  title        = {{The Agentic AI Security Scoping Matrix: A framework for securing autonomous AI systems}},
  howpublished = {AWS Security Blog},
  year         = {2025},
  url          = {https://aws.amazon.com/cn/blogs/security/the-agentic-ai-security-scoping-matrix-a-framework-for-securing-autonomous-ai-systems/},
  note         = {Published: 21 Nov 2025. Accessed: 29 Dec 2025}
}

@article{2002_A_taxonomy_of_web_search,
author = {Broder, Andrei},
title = {A Taxonomy of Web Search},
year = {2002},
journal = {SIGIR Forum},
}

@inproceedings{2004_Understanding_user_goals_in_web_search,
author = {Rose, Daniel E. and Levinson, Danny},
title = {Understanding User Goals in Web Search},
year = {2004},
booktitle = {International Conference on World Wide Web (WWW)},
}

@Article{2011_Query_reformulation_mining,
author={Boldi, Paolo
and Bonchi, Francesco
and Castillo, Carlos
and Vigna, Sebastiano},
title={Query reformulation mining: models, patterns, and applications},
journal={Information Retrieval},
year={2011},
}

@inproceedings{2009_Analyzing_and_evaluating_query_reformulation,
author = {Huang, Jeff and Efthimiadis, Efthimis N.},
title = {{Analyzing and evaluating query reformulation strategies in web search logs}},
year = {2009},
booktitle = {Conference on Information and Knowledge Management (CIKM)},
}

@article{2016_Towards_Searching_as_Learning,
  author  = {Rieh, Soo Young and Collins-Thompson, Kevyn and Hansen, Preben and Lee, Hye-Jung},
  title   = {Towards searching as a learning process: A review of current perspectives and future directions},
  journal = {Journal of Information Science},
  year    = {2016},
}

@article{2022_Learning_assessments_in_search_as_learning,
title = {Learning assessments in search-as-learning: A survey of prior work and opportunities for future research},
journal = {Information Processing and Management},
year = {2022},
author = {Kelsey Urgo and Jaime Arguello},
}

@inproceedings{2015_An_Eye_Tracking_Study_of_Query_Reformulation,
author = {Eickhoff, Carsten and Dungs, Sebastian and Tran, Vu},
title = {{An Eye-Tracking Study of Query Reformulation}},
year = {2015},
booktitle = {International Conference on Research and Development in Information Retrieval (SIGIR)},
}

@misc{2025_openai_gpt5_nano_model_docs,
  author       = {{OpenAI}},
  title        = {{GPT-5 nano Model}},
  howpublished = {OpenAI API Documentation},
  year         = {2025},
  url          = {https://platform.openai.com/docs/models/gpt-5-nano},
  note         = {Accessed: 2025-12-29}
}

@misc{2025_qwen3embedding,
  title={{Qwen3 Embedding: Advancing Text Embedding and Reranking Through Foundation Models}},
  author={Zhang, Yanzhao and Li, Mingxin and Long, Dingkun and Zhang, Xin and Lin, Huan and Yang, Baosong and Xie, Pengjun and Yang, An and Liu, Dayiheng and Lin, Junyang and Huang, Fei and Zhou, Jingren},
  year={2025},
  eprint={2506.05176},
  archivePrefix={arXiv}
}

@inproceedings{2007_Information_re_retrieval,
    author = {Teevan, Jaime and Adar, Eytan and Jones, Rosie and Potts, Michael A. S.},
    title = {Information re-retrieval: repeat queries in Yahoo's logs},
    year = {2007},
    booktitle = {International Conference on Research and Development in Information Retrieval (SIGIR)},
}

@inproceedings{2024_Adaptive_RAG,
    title = "{{Adaptive-{RAG}: Learning to Adapt Retrieval-Augmented Large Language Models through Question Complexity}}",
    author = "Jeong, Soyeong  and
      Baek, Jinheon  and
      Cho, Sukmin  and
      Hwang, Sung Ju  and
      Park, Jong",
    booktitle = "Conference of the North American Chapter of the Association for Computational Linguistics (NAACL)",
    year = "2024",

}

@inproceedings{2023_IR_COT,
    title = "{{Interleaving Retrieval with Chain-of-Thought Reasoning for Knowledge-Intensive Multi-Step Questions}}",
    author = "Trivedi, Harsh  and
      Balasubramanian, Niranjan  and
      Khot, Tushar  and
      Sabharwal, Ashish",
    booktitle = "Annual Meeting of the Association for Computational Linguistics (ACL)",
    year = "2023",
}

@article{2023_Lost_in_the_middle,
    title = "{{Lost in the Middle: How Language Models Use Long Contexts}}",
    author = "Liu, Nelson F.  and
      Lin, Kevin  and
      Hewitt, John  and
      Paranjape, Ashwin  and
      Bevilacqua, Michele  and
      Petroni, Fabio  and
      Liang, Percy",
    journal = "Transactions of the Association for Computational Linguistics",
    year = "2024",
}

@inproceedings{
    levy2025stwebagentbench,
    title={{{ST}-WebAgentBench: A Benchmark for Evaluating Safety and Trustworthiness in Web Agents}},
    author={Ido Levy and Ben wiesel and Sami Marreed and Alon Oved and Avi Yaeli and Segev Shlomov},
    booktitle={ICML Workshop on Computer Use Agents (ICML WCUA)},
    year={2025},
}

@misc{li2024llmsasjudgescomprehensivesurveyllmbased,
      title={{LLMs-as-Judges: A Comprehensive Survey on LLM-based Evaluation Methods}}, 
      author={Haitao Li and Qian Dong and Junjie Chen and Huixue Su and Yujia Zhou and Qingyao Ai and Ziyi Ye and Yiqun Liu},
      year={2024},
      eprint={2412.05579},
      archivePrefix={arXiv},
}

@inproceedings{
    asai2024selfrag,
    title={{Self-{RAG}: Learning to Retrieve, Generate, and Critique through Self-Reflection}},
    author={Akari Asai and Zeqiu Wu and Yizhong Wang and Avirup Sil and Hannaneh Hajishirzi},
    booktitle={International Conference on Learning Representations (ICLR)},
    year={2024},
}

@misc{wang2025aiagentshumanwork,
      title={{How Do AI Agents Do Human Work? Comparing AI and Human Workflows Across Diverse Occupations}}, 
      author={Zora Zhiruo Wang and Yijia Shao and Omar Shaikh and Daniel Fried and Graham Neubig and Diyi Yang},
      year={2025},
      eprint={2510.22780},
      archivePrefix={arXiv},
}

@misc{zhou2025psychologicalbehaviouralresponseshumanagent,
      title={Psychological and behavioural responses in human-agent vs. human-human interactions: a systematic review and meta-analysis}, 
      author={Jianan Zhou and Fleur Corbett and Joori Byun and Talya Porat and Nejra van Zalk},
      year={2025},
      eprint={2509.21542},
      archivePrefix={arXiv},
}

@inproceedings{feild2010predicting,
  title     = {{Predicting Searcher Frustration}},
  author    = {Feild, Henry A. and Allan, James and Jones, Rosie},
  booktitle = {International Conference on Research and Development in Information Retrieval (SIGIR)},
  year      = {2010},
}

@misc{2025_FlashResearch,
      title={{FlashResearch: Real-time Agent Orchestration for Efficient Deep Research}}, 
      author={Lunyiu Nie and Nedim Lipka and Ryan A. Rossi and Swarat Chaudhuri},
      year={2025},
      eprint={2510.05145},
      archivePrefix={arXiv},
}

@misc{2025_deep_research_gym,
      title={{DeepResearchGym: A Free, Transparent, and Reproducible Evaluation Sandbox for Deep Research}}, 
      author={João Coelho and Jingjie Ning and Jingyuan He and Kangrui Mao and Abhijay Paladugu and Pranav Setlur and Jiahe Jin and Jamie Callan and João Magalhães and Bruno Martins and Chenyan Xiong},
      year={2025},
      eprint={2505.19253},
      archivePrefix={arXiv},
}

@misc{2024_chatbot_arena,
  title   = {{Chatbot Arena: An Open Platform for Evaluating {LLM}s by Human Preference}},
  author  = {Chiang, Wei-Lin and Zheng, Lianmin and Sheng, Ying and Angelopoulos, Anastasios Nikolas and Li, Tianle and Li, Dacheng and Zhang, Hao and Zhu, Banghua and Jordan, Michael and Gonzalez, Joseph E. and Stoica, Ion},
  year    = {2024},
  eprint={2403.04132},
  archivePrefix={arXiv},
}

@inproceedings{2024_lmsys_chat_1m,
  title     = {{{LMSYS}-Chat-1M: A Large-Scale Real-World {LLM} Conversation Dataset}},
  author    = {Zheng, Lianmin and Chiang, Wei-Lin and Sheng, Ying and Li, Tianle and Zhuang, Siyuan and Wu, Zhanghao and Zhuang, Yonghao and Li, Zhuohan and Lin, Zi and Xing, Eric P. and Gonzalez, Joseph E. and Stoica, Ion and Zhang, Hao},
  booktitle = {International Conference on Learning Representations (ICLR)},
  year      = {2024},
}

@techreport{2025_how_people_use_chatgpt,
  title       = {{How People Use {ChatGPT}}},
  author      = {{OpenAI}},
  year        = {2025},
  institution = {OpenAI},
}

@misc{2025_economic_tasks_ai_claude,
  title   = {{Which Economic Tasks are Performed with {AI}? Evidence from Millions of {Claude} Conversations}},
  author  = {Handa, Kunal and Tamkin, Alex and McCain, Miles and Huang, Saffron and Durmus, Esin and Heck, Sarah and Mueller, Jared and Hong, Jerry and Ritchie, Stuart and Belonax, Tim and Troy, Kevin K. and Amodei, Dario and Kaplan, Jared and Clark, Jack and Ganguli, Deep},
  year    = {2025},
  eprint={2503.04761},
  archivePrefix={arXiv},
}

@misc{2025_scierena,
  title   = {{SciArena: An Open Evaluation Platform for Foundation Models in Scientific Literature Tasks}},
  author  = {Zhao, Yilun and Zhang, Kaiyan and Hu, Tiansheng and Wu, Sihong and Le Bras, Ronan and Anderson, Taira and Bragg, Jonathan and Chang, Joseph Chee and Dodge, Jesse and Latzke, Matt and Liu, Yixin and McGrady, Charles and Tang, Xiangru and Wang, Zihang and Zhao, Chen and Hajishirzi, Hannaneh and Downey, Doug and Cohan, Arman},
  year    = {2025},
  eprint={2507.01001},
  archivePrefix={arXiv},
}

@misc{2023_webarena,
  title   = {WebArena: A Realistic Web Environment for Building Autonomous Agents},
  author  = {Zhou, Shuyan and Xu, Frank F. and Zhu, Hao and Zhou, Xuhui and Lo, Robert and Sridhar, Abishek and Cheng, Xianyi and Ou, Tianyue and Bisk, Yonatan and Fried, Daniel and Alon, Uri and Neubig, Graham},
  year    = {2023},
  eprint={2307.13854},
  archivePrefix={arXiv},
}

@inproceedings{2022_webshop,
  title     = {{WebShop: Towards Scalable Real-World Web Interaction with Grounded Language Agents}},
  author    = {Yao, Shunyu and Chen, Howard and Yang, John and Narasimhan, Karthik},
  booktitle = {International Conference on Neural Information Processing Systems (NeurIPS)},
  year      = {2022},
}

@misc{2023_agentbench,
  title   = {{AgentBench: Evaluating {LLM}s as Agents}},
  author  = {Liu, Xiao and Yu, Hao and Zhang, Hanchen and Xu, Yifan and Lei, Xuanyu and Lai, Hanyu and Gu, Yu and Ding, Hangliang and Men, Kaiwen and Yang, Kejuan and Zhang, Shudan and Deng, Xiang and Zeng, Aohan and Du, Zhengxiao and Zhang, Chenhui and Shen, Sheng and Zhang, Tianjun and Su, Yu and Sun, Huan and Huang, Minlie and Dong, Yuxiao and Tang, Jie},
  year    = {2023},
  eprint={2308.03688},
  archivePrefix={arXiv}
}

@misc{2023_toolllm,
  title   = {{ToolLLM}: Facilitating Large Language Models to Master 16,000+ Real-world {APIs}},
  author  = {Qin, Yujia and Liang, Weitian and Yang, Jiarui and Zhou, Bohan and Yan, Zonghao and Lu, Weixuan and Liu, Qian and Hu, Shengding and Huang, Yufan and Zeng, Yijia and others},
  year    = {2023},
  eprint={2307.16789},
  archivePrefix={arXiv}
}

@misc{2025_gemini_3_flash,
  author       = {{Google}},
  title        = {{Gemini 3 Developer Guide} (Model ID: gemini-3-flash-preview)},
  howpublished = {Google AI for Developers Documentation},
  year         = {2025},
  note         = {Gemini 3 models in preview; model IDs listed in documentation},
  url          = {https://ai.google.dev/gemini-api/docs/gemini-3},
  urldate      = {2026-01-18}
}

@misc{2025_hle,
  title   = {{Humanity's Last Exam}},
  author  = {Phan, Long and Gatti, Alice and Han, Ziwen and others},
  year    = {2025},
  eprint={2501.14249},
  archivePrefix={arXiv}
}

@inproceedings{frames_dataset,
  author       = {Satyapriya Krishna and
                  Kalpesh Krishna and
                  Anhad Mohananey and
                  Steven Schwarcz and
                  Adam Stambler and
                  Shyam Upadhyay and
                  Manaal Faruqui},
  title        = {{Fact, Fetch, and Reason: {A} Unified Evaluation of Retrieval-Augmented
                  Generatio}},
  booktitle    = {Conference of the Nations of the Americas
                  Chapter of the Association for Computational Linguistics (NAACL)},
  year         = {2025},
}

@inproceedings{webwalker_dataset,
  author       = {Jialong Wu and
                  Wenbiao Yin and
                  Yong Jiang and
                  Zhenglin Wang and
                  Zekun Xi and
                  Runnan Fang and
                  Linhai Zhang and
                  Yulan He and
                  Deyu Zhou and
                  Pengjun Xie and
                  Fei Huang},
  title        = {{WebWalker: Benchmarking LLMs in Web Traversa}},
  booktitle    = {Annual Meeting of the Association for Computational
                  Linguistics},

  year         = {2025},

}

@inproceedings{DBLP:conf/ercimdl/GomesMC19,
  author       = {Pedro Gomes and
                  Bruno Martins and
                  Lu{\'{\i}}s Cruz},
  title        = {{Segmenting User Sessions in Search Engine Query Logs Leveraging Word
                  Embeddings}},
  booktitle    = {International Conference
                  on Theory and Practice of Digital Libraries (TPDL)},
  year         = {2019}
}
